\documentclass[preprint,11pt,authoryear]{elsarticle}
\usepackage[top=2.5cm,bottom=2.5cm,left=3cm,right=3cm]{geometry}

\usepackage{natbib}
\usepackage{hyperref}
\makeatletter
\renewcommand\hyper@natlinkbreak[2]{#1}
\makeatother
\usepackage{etoolbox}
\usepackage{tcolorbox}
\usepackage{graphicx}
\setcitestyle{aysep={}} 
\usepackage{amsmath}
\usepackage[hang]{footmisc}
\usepackage{lipsum}
\usepackage{mathrsfs}
\usepackage{bbm}
\usepackage{indentfirst} 
\usepackage{enumerate}
\usepackage{microtype}
\usepackage{geometry}
\usepackage{subfigure}
\usepackage{amssymb}
\usepackage[misc]{ifsym}
\usepackage{mathtools}
\usepackage{soul}
\setstcolor{blue}
\usepackage[pagewise]{lineno}
\usepackage{pdflscape}
\usepackage{setspace}
\usepackage[linesnumbered,ruled,vlined]{algorithm2e}
\usepackage{xcolor}
\usepackage{lmodern} 
\usepackage{graphicx,scalerel}
\usepackage{nameref,hyperref}
\usepackage{caption}
\usepackage{makecell, multirow}
\setcellgapes{3pt}
\usepackage{multirow}
\usepackage{adjustbox}
\usepackage{booktabs}
\usepackage{afterpage}
\usepackage[compact]{titlesec}

\newcommand{\Gorder}[1]{}

\newenvironment{stretchpars}
 {\par\setlength{\parfillskip}{0pt}}
 {\par}

\newcommand\newbar[1]{\hstretch{2}{\bar{\hstretch{.5}{#1\mkern2mu}}}\mkern-1mu}

\journal{arXiv}
\begin{document}
\begin{frontmatter}

\title{Interpreting how nonlinear diffusion affects the fate of bistable populations using a discrete modelling framework}

\author{Yifei Li$^1$,  Pascal R. Buenzli$^1$, Matthew J. Simpson$^1$}

\address{$^1$School of Mathematical Sciences, Queensland University of  Technology, Brisbane, Australia.}

\begin{abstract}
Understanding whether a population will survive and flourish or become extinct is a central question in population biology.  One way of exploring this question is to study population dynamics using reaction-diffusion equations, where migration is usually represented as a linear diffusion term, and birth-death is represented with a bistable source term.  While linear diffusion is most commonly employed to study migration, there are several limitations of this approach, such as the inability of linear diffusion-based models to predict a well-defined population front.  One way to overcome this is to generalise the constant diffusivity, $D$, to a nonlinear diffusivity function $D(C)$, where $C>0$ is the density.  While it has been formally established that the choice of $D(C)$ affects long-term survival or extinction of a bistable population, working solely in a classical continuum framework makes it difficult to understand precisely how the choice of $D(C)$ affects survival or extinction.  Here, we address this question by working with a simple discrete simulation model that is easy to interpret.  The continuum limit of the discrete model is a nonlinear reaction-diffusion equation, where the flux involves a nonlinear diffusion term and the source term is given by the strong Allee effect bistable model.  We study population extinction/survival using this very intuitive discrete framework together with numerical solutions of the reaction-diffusion continuum limit equation.  This approach provides clear insight into how the choice of $D(C)$ either encourages or suppresses population extinction relative to the classical linear diffusion model.
\end{abstract}


\end{frontmatter}


\newpage

\section{Introduction}
\label{sec_1}
Predicting whether a population will survive or go extinct is a key question in population biology \citep{Joel1990,Robert1998,Axelrod2006,otso2010,Sonia2011}. For example, predicting whether a species released into a wild area will survive is crucial in protecting endangered animals \citep{Saltz1995}. Similarly, whether cancer spreads to a different body part from a primary tumour site depends on the survival of small numbers of tumour cells growing successfully in new locations \citep{Korolev2014,Gerlee2021}. A classical continuum model for studying the survival of biological populations is the strong Allee effect model, based on an ordinary differential equation (ODE),
\begin{equation}
    \label{Intro_ODE}
    \frac{\textls{d}C(t)}{\textls{d}t}=\lambda C(t)\left(1-\frac{C(t)}{K}\right)\left(\frac{C(t)}{A}-1\right),
\end{equation}
where $C(t)\ge0$ is the population density at time $t\ge0$, $\lambda>0$ is the intrinsic growth rate, $K>0$ is the carrying capacity density, and $0<A<K$ is the Allee threshold density \citep{Allee1932,Kot2001,Keshet2005,courchamp2008,Taylor2015Allee,Anudeep2020,Fadai2020}. The fate of a population described by \eqref{Intro_ODE} depends solely upon the initial density, $C(0)$. Extinction occurs if $C(0) < A$, leading to $C(t) \to 0$ as $t \to \infty$. In contrast, the population survives if $C(0) > A$, leading to $C(t) \to K$ as $t \to \infty$. Such
dynamics, leading to either eventual survival or extinction, are sometimes called \textit{bistable} population dynamics. The strong Allee effect model belongs to a broader class of population models, called bistable population dynamics models, and the key feature of these models is that they involve three equilibrium points: $C=0$ and $C=K>0$ are stable equilibrium points, and $C = A$, where $0< A< K$ is unstable. There are many ODE models of this kind, not just the classical cubic form in \eqref{Intro_ODE} \citep{Johnson2019,Fadai2020BMB,Alkhayuon2021,Gerlee2021}.

To investigate spatial effects, such as moving invasion fronts, some studies consider incorporating Equation \eqref{Intro_ODE} into a reaction-diffusion equation, where the population density depends upon both position and time \citep{hadeler1975travelling,lewis1993allee,holmes1994,Hastings2005,Gabriel2015,johnston2017co,Neufel2017,Maud2019,Li2020travel,Li2020shock}. In reaction-diffusion models, the dynamics of bistable populations involve a more complicated interaction between the bistable source term and the diffusion term. Unlike ODE models where the fate of a bistable population is solely determined by the initial density, many factors influence whether the population will survive or go extinct in reaction-diffusion models \citep{lewis1993allee,BRADFORD1970a,BRADFORD1970b,Johnston2020,li2021dimensionality}. For example, the initial area of a bistable population on an infinite domain needs to be greater than a threshold, called the \textit{critical initial area}, so that the population avoids extinction \citep{lewis1993allee}.

Most reaction-diffusion models in population biology consider a constant diffusivity associated with Fick's first law of diffusion, which states that the diffusive flux is proportional to the spatial gradient of density \citep{hadeler1975travelling,holmes1994,Kot2001,Hastings2005,Jin2016Sto,murray2002mathematical,Gabriel2015,Neufel2017}. In one spatial dimension the diffusive flux is $J=-D\partial C(x,t)/\partial x$, where $D>0$ is the constant diffusivity. Linear diffusion is popular in modelling biological populations, since this model is very simple, and has a straightforward connection with a range of underlying stochastic models, such as conceptualising the motion of individuals in the population as a simple unbiased random walk in the dilute limit, where interactions between individuals are weak \citep{hughes1995random,Kot2001,murray2002mathematical,liggett2013stochastic}. However, despite the immense popularity of linear diffusion, there are well-documented circumstances where population dynamics cannot be described by this simple model. For example, sharp moving fronts in cell migration assays cannot be represented by linear diffusion, and so there has been great interest in modelling the motion of well-defined fronts using degenerate nonlinear diffusion terms \citep{maini2004travelling,maini2004wound,Sengers2007,Jin2016reproducibility,Scott2019Hole}. Similar to cell biology applications, mathematical models of insect dispersal with linear diffusion are unable to replicate observations where well-defined sharp fronts play an important role \citep{Shigesada1980,murray2002mathematical}. Therefore, reaction-diffusion equations with nonlinear diffusion are considered in a variety of applications where, in one spatial dimension, the flux is $J=-D(C(x,t))\partial C(x,t)/\partial x$, with the key difference that the constant linear diffusivity $D$ is now generalised to a nonlinear function $D(C)>0$ \citep{Shigesada1980,murray2002mathematical,Painter2003,maini2004travelling,Sengers2007,Cai2007,deroulers2009modeling,fernando2010nonlinear,johnston2012mean,Martinez2015,yates2015incorporating,Jin2016reproducibility,Scott2019Hole,Bubba2020}. Unlike the constant diffusivity that can be interpreted as undirected random motion of individuals without interaction \citep{hughes1995random,liggett2013stochastic}, identifying the behaviour of individuals corresponding to a given nonlinear diffusion term is less clear. Therefore, it is not always obvious which nonlinear diffusion term is appropriate to model a particular situation \citep{Sherratt1990,murray2002mathematical,Cai2007,Jin2016reproducibility}. Since
it is known that nonlinear diffusion can impact the conditions required for survival of a bistable population \citep{cantrell2004spatial,lee2006bifurcation}, exploring the behaviour of individuals is helpful to provide a simple interpretation of how $D(C)$ affects the fate of bistable populations subject to a nonlinear diffusion migration mechanism. Therefore, it is valuable to study the connection between the behaviour of individuals and the nonlinear diffusion term in reaction-diffusion equations.

To connect continuum models with the behaviour of individuals, we work with a physically intuitive discrete framework. The discrete model incorporates straightforward crowding effects into birth, death and movement of individuals on a two-dimensional hexagonal lattice \citep{li2021dimensionality}. In particular, we quantify the influence of crowdedness on the motility of individuals by using a crowding function $G(C)>0$, which explicitly describes how the local crowding affects the ability of individuals to move. The continuum limit of the discrete model is a reaction-diffusion equation with a strong Allee effect source term, and a general nonlinear diffusivity function $D(C)$. This framework allows us to investigate population dynamics through repeated simulation of the discrete model, as well as solving the associated reaction-diffusion continuum limit model numerically. Through the mathematical relationship between the nonlinear diffusivity function $D(C)$ and the underlying crowding function $G(C)$, we develop an intuitive understanding of how different choices of $D(C)$ affect the extinction or survival of the population. To improve our understanding, we derive expressions for the density-dependent flux of populations associated with the discrete model. The expression for the flux can be re-written as the flux associated with a linear diffusion mechanism plus a term, which we interpret as a correction that is associated with the effects of nonlinear diffusion. Writing the flux in this way allows us to directly relate how different choices of $D(C)$ either encourage or suppress extinction. All interpretations of our modelling framework are supported by a suite of stochastic simulations and numerical solutions of the associated continuum limit reaction-diffusion equation.  All numerical algorithms required to replicate our work are available on \href{https://github.com/oneflyli/YifeiNonliearDiffusion2021}{Github}.

\section{The discrete model and the continuum limit}
\label{sec_2}
In this section we introduce a lattice-based discrete model and the corresponding continuum limit model description that is closely related to our previous work \citep{li2021dimensionality}. Unlike the work in \citet{li2021dimensionality}, which only considers examples where the motility of individuals is given by a linear diffusion mechanism, here we focus on a more broad range of motility mechanisms that include a range of choices of nonlinear diffusivity functions.

In the discrete model individuals are represented as agents on a two-dimensional hexagonal lattice with spacing $\Delta>0$. A lattice site $\mathbf{s}$, indexed by $(i,j)$ with a unique Cartesian coordinate $(x,y)$, is either occupied $C_\mathbf{s}=1$, or vacant $C_\mathbf{s}=0$. Stochastic simulations include birth, death and movement events, and we will now explain the details of these mechanisms.

If there are $Q(t)$ agents on the lattice, we use a random sequential updating method to evolve the discrete model from time $t$ to time $t + \tau$.  To achieve this we select $Q(t)$ agents at random, with replacement, and give those agents an opportunity to undergo a movement event.  We then select another $Q(t)$ agents, at random, with replacement, and give those agents an opportunity to undergo a birth/death event. Once these two sets of events have been assessed, we advance time from $t$ to $t + \tau$ and repeat until the desired output time is reached \citep{simpson2010cell}. 

\begin{figure}
\centering
\includegraphics[width=0.75\textwidth]{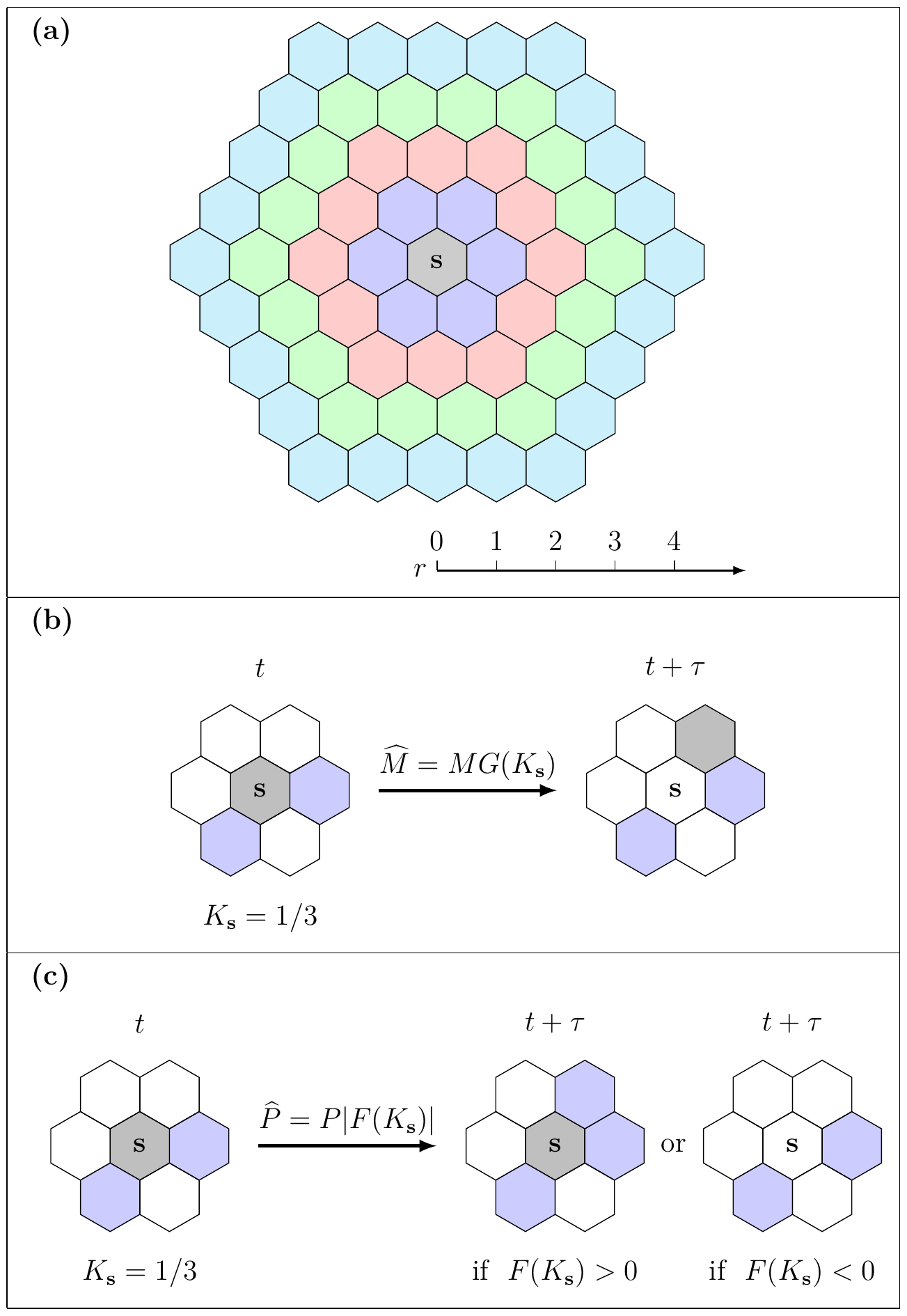}
\caption{\textbf{Movement and growth mechanisms.} (a) Different-sized spatial templates $\mathcal{N}_r$, where $r$ is the diameter of the concentric rings surrounding site $\mathbf{s}$. (b) Movement mechanism with $r=1$. (c) Growth mechanism with $r=1$. In (b) and (c), there are two neighbouring agents (blue) surrounding the agent at site $\mathbf{s}$ (grey) at time $t$. The agent at site $\mathbf{s}$ undergoes a movement event with probability $\widehat{M}=MG(K_\mathbf{s})$, where $K_\mathbf{s}=1/3$, as shown in (b). As there are four vacant neighbouring sites, the probability of moving to one of the vacant sites is $\widehat{M}/4$. Similarly, the agent undergoes a growth event with probability $\widehat{P}=P\lvert F(K_\mathbf{s})\lvert$, where $K_\mathbf{s}=1/3$, as shown in (c). If $F(K_\mathbf{s})>0$, the potential growth event is a birth event. As there are four vacant neighbouring sites, the probability of placing a new agent in one of the vacant sites is $\widehat{P}/4$. If $F(K_\mathbf{s})<0$, the potential growth event is a death event. That is, the agent will be removed out of the lattice with probability $\widehat{P}$.} 
\label{fig:hex} 
\end{figure}

For a potential motility event, if the agent in question is at site $\mathbf{s}$, that agent will move with probability $\widehat{M} = MG(K_\mathbf{s})$, where $M$ is the probability that an isolated agent will attempt to move during a time interval of duration $\tau$, and $G(K_\mathbf{s}) \in [0,1]$ is a \textit{movement crowding function} which quantities how crowding in a small neighbourhood of $\mathbf{s}$ influences motility. We interpret $G(K_\mathbf{s})$ to be a measure of the influence of the local density upon movement since $K_\mathbf{s}$ is a simple measure of density around site $\mathbf{s}$, given by 
\begin{equation}
    \label{localdensity}
    K_\mathbf{s}=\frac{1}{\lvert\mathcal{N}_r\lvert}\sum_{s'\in\mathcal{N}_r\{\mathbf{s}\}}C_{\mathbf{s}'}\in[0,1],
\end{equation}
where $\mathcal{N}_r\{\mathbf{s}\}$ denotes the set of neighbouring sites surrounding site $\mathbf{s}$. Since the local density can be measured with different-sized spatial templates, we use $r$ to represent the diameter of concentric rings surrounding site $\mathbf{s}$, so that the number of neighbouring sites of any site is $\lvert\mathcal{N}_r\lvert=3r(r+1)$, as shown in Figure~\ref{fig:hex}(a).

If a movement event occurs, the agent at site $\mathbf{s}$ will move into a randomly chosen vacant site in $\mathcal{N}_r\{\mathbf{s}\}$. Therefore, the probability for the agent at site $\mathbf{s}$ moving to one of the vacant sites is $\widehat{M}/(\lvert\mathcal{N}_r\lvert(1-K_\mathbf{s}))$.  We show the movement mechanism with $r=1$ leading to $\lvert\mathcal{N}_r\lvert=6$ in Figure~\ref{fig:hex}(b). In this particular configuration the agent at site $\mathbf{s}$ has two neighbouring agents, giving $K_\mathbf{s}=1/3$. The probability of undergoing a movement event is $\widehat{M}=MG(1/3)$. As there are four vacant sites in $\mathcal{N}_1\{\mathbf{s}\}$, the probability of moving to one of the vacant sites is $\widehat{M}/4$. Note that we require $G(1)=0$ as individuals have no space to move if their neighbouring sites are all occupied.

For a potential growth event, if the agent in question is at site $\mathbf{s}$, that agent will undergo a growth event with probability $\widehat{P}= PF(K_\mathbf{s})$, where $P$ is the probability that an isolated agent will attempt to undergo a growth event during a time interval of duration $\tau$, and $F(K_\mathbf{s}) \in [-1,1]$ is the \textit{growth  crowding function} which quantities how crowding in a small neighbourhood of $\mathbf{s}$ influences the propensity of agents to proliferate or die.  Since the growth mechanism includes both proliferation and death as potential outcomes, we define $F(K_\mathbf{s}) \in [-1,1]$ such that a proliferation event takes place when $F(K_\mathbf{s})>0$ and a death event takes place when $F(K_\mathbf{s})<0$.  In the case of a proliferation event, a new daughter agent will be placed on a randomly chosen vacant site in $\mathcal{N}_r(\mathbf{s})$, whereas if a death event takes place the agent at site $\mathbf{s}$ will be removed from the simulation.   After $Q(t)$ potential growth events have been assessed, the value of $Q(t)$ is updated accordingly.  

We show the growth mechanism with $r=1$ in Figure~\ref{fig:hex}(c). As the agent at site $\mathbf{s}$ has two neighbouring agents, we have $K_\mathbf{s}=1/3$. Therefore, the probability of undergoing a growth event is $\widehat{P}=P\lvert F(1/3)\lvert$. If $F(1/3)>0$, as there are four vacant sites in $\mathcal{N}_1\{\mathbf{s}\}$, the agent will place a new agent on one of the vacant sites with probability $\widehat{P}/4$. If $ F(1/3)<0$, the agent will be removed from the lattice with probability $\widehat{P}$. 

\begin{landscape}
\begin{figure}
\centering
\includegraphics[width=1.5\textwidth]{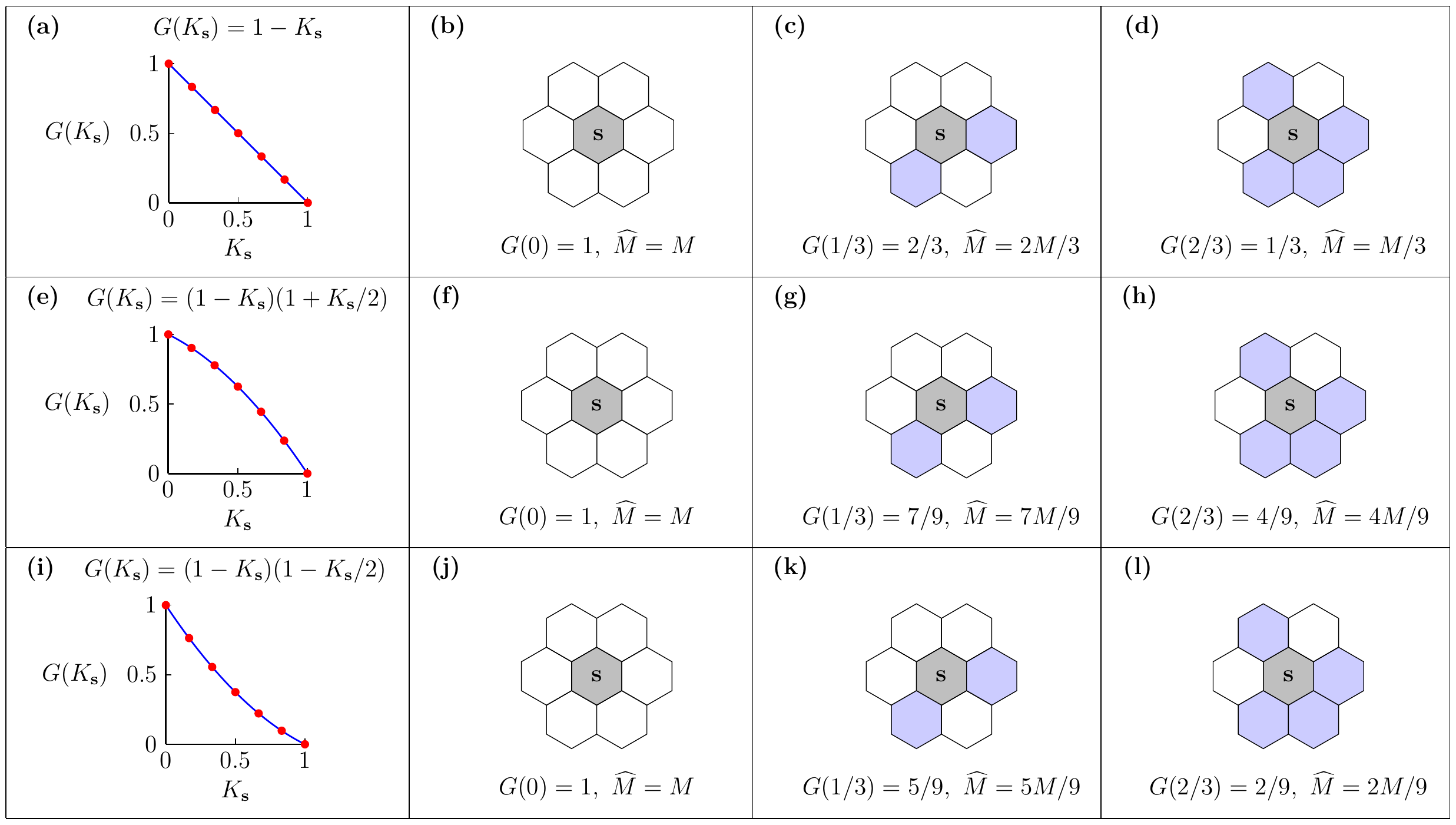}
\caption{\textbf{Movement mechanism with different movement crowding functions on a spatial template with $r=1$.} In each subfigure we represent the agent at site $\mathbf{s}$ with grey, the neighbouring agents with blue, and the vacant neighbouring sites with white. (a) $G(K_\mathbf{s})=1-K_\mathbf{s}$. The red dots represent the values of $G(K_\mathbf{s})$ that can be measured from the discrete model. (b)--(d) Probabilities of undergoing a movement event for the agent at site $\mathbf{s}$ associated with the $G(K_\mathbf{s})$ in (a). (e) $G(K_\mathbf{s})=(1-K_\mathbf{s})(1+K_\mathbf{s}/2)$. (f)--(h) Probabilities of undergoing a movement event for the agent at site $\mathbf{s}$ associated with the $G(K_\mathbf{s})$ in (e). (i) $G(K_\mathbf{s})=(1-K_\mathbf{s})(1-K_\mathbf{s}/2)$. (j)--(l) Probabilities of undergoing a movement event for the agent at site $\mathbf{s}$ associated with the $G(K_\mathbf{s})$ in (i).} 
\label{fig:hex2} 
\end{figure}
\end{landscape}

\begin{stretchpars}
A key feature of the discrete model is that we use $G(K_\mathbf{s})$ to explicitly describe the influence of crowding effects on the movement of individuals. We provide several examples of $G(K_\mathbf{s})$ and show how they influence the movement of agents on the spatial template with $r=1$ in Figure~\ref{fig:hex2}. We first consider $G(K_\mathbf{s})=1-K_\mathbf{s}$ which has a constant slope, as shown in Figure~\ref{fig:hex2}(a). The probability of a movement event with $K_\mathbf{s}=0$, $K_\mathbf{s}=1/3$, and $K_\mathbf{s}=2/3$ is given in Figures~\ref{fig:hex2}(b)--(d), respectively. With this simple movement crowding function, the probability of the agent at site $\mathbf{s}$ moving to one of its neighbouring vacant sites is always $M/6$, which is independent of the local density, $K_\mathbf{s}$. We then consider a concave down function, $G(K_\mathbf{s})=(1-K_\mathbf{s})(1+K_\mathbf{s}/2)$ in Figure~\ref{fig:hex2}(e). The probability of a movement event with $K_\mathbf{s}=0$, $K_\mathbf{s}=1/3$, and $K_\mathbf{s}=2/3$ is given in Figures~\ref{fig:hex2}(f)--(h), respectively. Compared to the simplest crowding function $G(K_\mathbf{s})=1-K_\mathbf{s}$, the agent has a larger net movement probability for $K_\mathbf{s}\in(0,1)$. Finally, we consider a concave up function, $G(K_\mathbf{s})=(1-K_\mathbf{s})(1-K_\mathbf{s}/2)$ in Figure~\ref{fig:hex2}(i). The probability of a movement event with $K_\mathbf{s}=0$, $K_\mathbf{s}=1/3$, and $K_\mathbf{s}=2/3$ is shown in Figures~\ref{fig:hex2}(j)--(l), respectively. In this case, the agent has a reduced probability of movement for $K_\mathbf{s}\in(0,1)$, relative to the simplest case $G(K_\mathbf{s})=1-K_\mathbf{s}$. The growth mechanism of agents applies a similar way of incorporating the influence of crowding effects
\end{stretchpars}

\noindent
into the discrete model \citep{li2021dimensionality}. We present the pseudo-code for implementing a single realisation of the discrete model in the Supplementary Material document.

If we consider the spatial template with $r=1$ for the movement mechanism and $r\ge1$ for the growth mechanism of agents, the continuum limit of the discrete model is
\begin{equation}
    \label{PDE_general_2D}
    \frac{\partial C(x,y,t)}{\partial t}=\nabla\cdot\left(D(C(x,y,t))\nabla C(x,y,t)\right)+R(C),
\end{equation}
with nonlinear diffusivity function
\begin{equation}
\label{relation}
    D(C)=D_0\left[C\frac{\textls{d} G(C)}{\textls{d}C}+\frac{1+C}{1-C}G(C)\right],
\end{equation}
and a source term $R(C)=\lambda CF(C)$, where
\begin{equation}
\label{limitparameter1}
    D_0=\lim_{\Delta,\tau\to0}
    \frac{M\Delta^2}{4\tau},\quad \lambda=\lim_{\tau\to0}\frac{P}{\tau}.
\end{equation}
Here, $D_0>0$ is a constant in the limit that $\Delta\to0$ and $\tau\to0$ with the ratio $\Delta^2/\tau$ held constant, and $\lambda>0$ is constant when $P=\mathcal{O}(\tau)$, which implies that the continuum limit is valid when $P\ll M$. Note that in the discrete model we have $K$ as the argument of the crowding function, and that in the continuum limit the argument of the crowding function is $C$. This difference can be reconciled through carrying out the full details of the discrete-to-continuum averaging arguments. Full algebraic details of the intermediate steps required to derive the continuum limit is given in the Supplementary Material. Throughout this study we work with dimensionless simulations by setting $\Delta=\tau=1$ in the discrete model, which leads to $D_0=M/4$ and $\lambda=P$ in the continuum limit. In cell biology, experimental observations imply that cell motility is reasonably well approximated by a nearest neighbour random walk whereas cell proliferation involves the disposition of daughter agents at some distance from the mother agent \citep{simpson2010cell}. Therefore, throughout this work we set $r=1$ for the motility mechanism and $r=4$ for the proliferation mechanism, which is consistent with previous modelling \citep{simpson2010cell,li2021dimensionality}. Moreover, as we are interested in the survival and extinction of populations, we choose a growth crowding function $F(C)=2.5 (1-C)(C-A)$ with $A=0.4$, which leads to a cubic source term, $R(C)$, associated with the strong Allee effect. With this choice of growth crowding function, we have $F(0)=-1$ indicating that isolated agents have the largest probability of dying. 

\section{Relationship between \textit{D(C)} and \textit{G(C)}}
\label{sec_3}
Based on Equation \eqref{relation}, we are now in a unique position where we can specify an intuitive crowding function for the discrete model, $G(C)$, and use the discrete-to-continuum framework to understand how this translates into a population-level nonlinear diffusivity function, $D(C)$. This approach is very different to the more usual approach of simply specifying some phenomenological $D(C)$ function, without any detailed understanding of how a particular choice of nonlinear diffusivity impacts the underlying discrete mechanism \citep{Sherratt1990,maini2004travelling,maini2004wound,Sengers2007,Scott2019Hole}. 

\begin{figure}
\centering
\includegraphics[width=\textwidth]{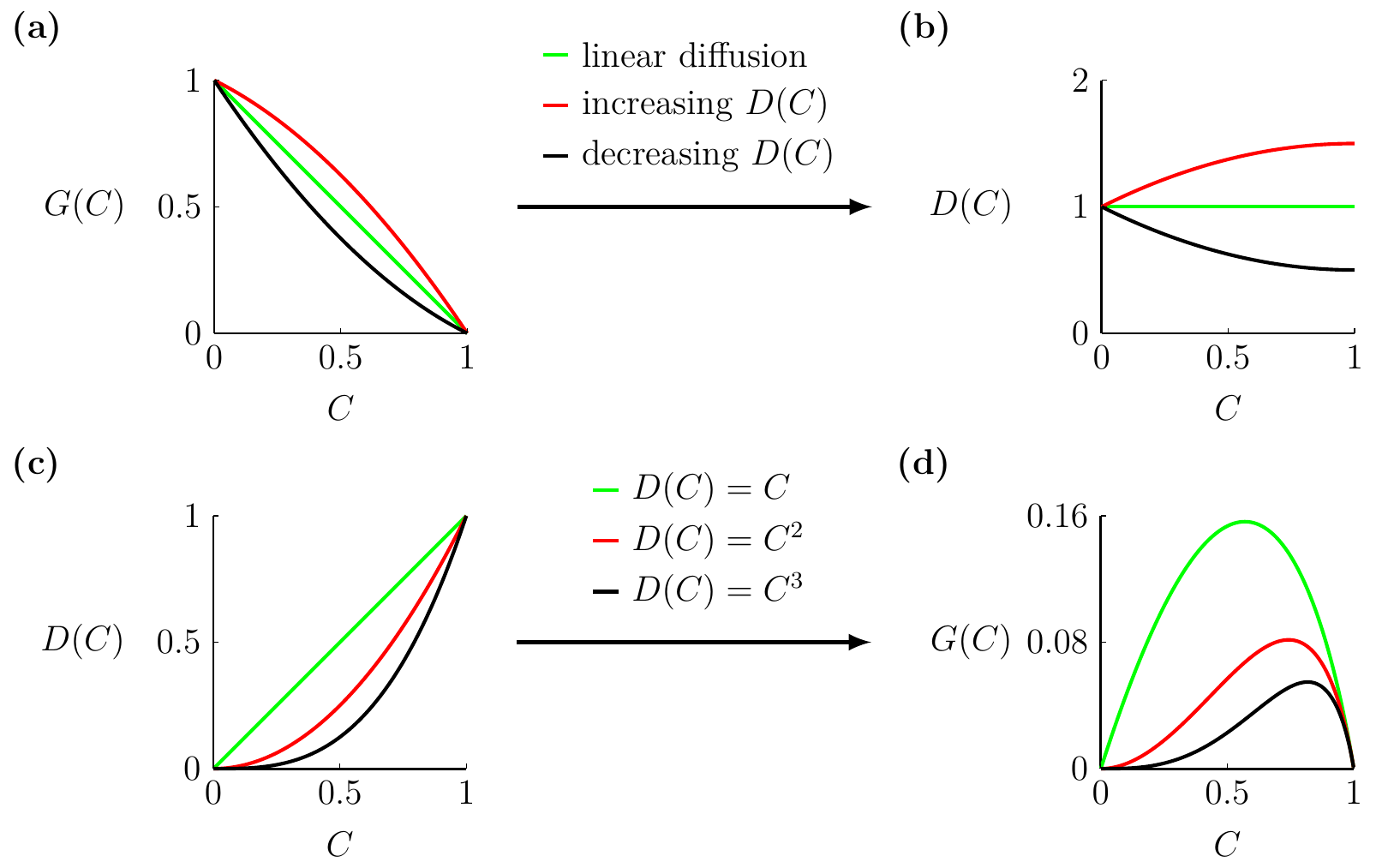}
\caption{\textbf{Two approaches of connecting the movement crowding function and the diffusivity function.} (a) Movement crowding functions $G(C)=1-C$ (green), $G(C)=(1-C)(1+C/2)$ (red) and $G(C)=(1-C)(1-C/2)$ (black). (b) Diffusivity functions associated with the movement crowding functions in (a) with $D_0=1$.
(c) Diffusivity functions $D(C)=D_0C$ (green), $D(C)=D_0C^2$ (red) and $D(C)=D_0C^3$ (black) with $D_0=1$. (d) Movement crowding functions associated with the nonlinear diffusivity functions in (c).} 
\label{fig:functions0} 
\end{figure}

There are two ways of taking advantage of \eqref{relation} to study population dynamics. First, for a given movement mechanism of individuals described by $G(C)$, we can simply substitute into this expression to give the corresponding $D(C)$. To demonstrate this first approach, we present three examples of $G(C)$, which were examined in Figure~\ref{fig:hex2}, and study the corresponding $D(C)$,
\begin{align}
    \label{eq1}
    &G(C)=1-C,\quad &&D(C)=D_0,\\
    \label{eq2}    
    &G(C)=(1-C)\left(1+\frac{C}{2}\right),\quad &&D(C)=D_0\left[1+C\left(1-\frac{C}{2}\right)\right],\\
    \label{eq3}    
    &G(C)=(1-C)\left(1-\frac{C}{2}\right),\quad &&D(C)=D_0\left[1-C\left(1-\frac{C}{2}\right)\right],
\end{align}
see Figures~\ref{fig:functions0}(a)--(b). In each of these three crowding functions we always have $G(0)=1$, which is reasonable since this condition implies that isolated agents are unaffected by crowding \citep{Jin2016Sto}. We first consider $G(C)=1-C$, which has a constant slope and leads to a constant diffusivity $D(C)=D_0$. As we mentioned in Section \ref{sec_2}, the probability of an agent moving to one of its neighbouring vacant sites is always $M/6$, which is independent of density. This is consistent with the continuum model where the standard linear diffusion mechanism means that the diffusivity is independent of density. Next we consider the concave down crowding function $G(C) = (1-C)(1+C/2)$, which has the property that $G(C) > 1-C$ for all $C \in (0,1)$.  For this crowding function we obtain an increasing nonlinear diffusivity function $D(C) > D_0$, which is reasonable since the motility of individuals is reduced less by crowding than in the case where $G(C) = 1-C$, corresponding to linear diffusion.  Similarly, the concave up crowding function $G(C) = (1-C)(1-C/2)$, which has the property that $G(C) < 1-C$ for all $C \in (0,1)$. For this crowding function we obtain a decreasing nonlinear diffusivity function $D(C)< D_0$, which again is reasonable since the motility of individuals is reduced more by crowding than in the case where $G(C)=1-C$, corresponding to linear diffusion.  

The first approach to use \eqref{relation} involves specifying a physically reasonable crowding function, $G(C)$, and using the discrete-to-continuum conservation argument to understand the corresponding population-level nonlinear diffusivity function, $D(C)$. Of course, we can also view \eqref{relation} as allowing us to specify a particular nonlinear diffusivity function $D(C)$, and to understand which particular crowding function is associated with that choice of $D(C)$.  One of the challenges associated with this second approach is that a given $D(C)$ may not lead to a physically realistic crowding function, as we will now explore. One of the most standard choices of nonlinear diffusivity is a power law diffusivity, $D(C) = D_0 C^m$, where $m$ is some constant exponent. This model has played a very important role in population biology models \citep{mat2011degenerate,Scott2019Hole}, since combining this nonlinear diffusivity term with a logistic source term gives rise to the well-studied Porous-Fisher model \citep{Sherratt1990,Witelski1995Fisher,murray2002mathematical,Jin2016reproducibility,buenzli2020}. If we consider $m \ge 0$ and $G(C) \in [0,1]$ we can solve \eqref{relation} to give
\begin{equation}
    \label{solveODE3}
    G(C)=\left(C^{m+2}-2C^{m+1}+C^{m}\right)\frac{~_2F_1(2,m+1;m+2;C)}{m+1},
\end{equation}
where $~_2F_1(2,m+1;m+2;C)$ is the hypergeometric function \citep{abramowitz1964handbook}. We present three examples with $m=1,2$ and $3$, given by
\begin{align}
    \label{eq4}
    &D(C)=D_0C,\quad &&G(C)=(1-C)\left(1+\frac{1-C}{C}\ln{(1-C)}\right),\\
    \label{eq5}    
    &D(C)=D_0C^2,\quad &&G(C)=(1-C)\left(2-C+\frac{2(1-C)}{C}\ln{(1-C)}\right),\\
    \label{eq6}    
    &D(C)=D_0C^3,\quad &&G(C)=(1-C)\left(\frac{-C^2-3C+6}{2}+\frac{3(1-C)}{C}\ln{(1-C)}\right).
\end{align}
Figures~\ref{fig:functions0}(c)--(d) compare the specified $D(C)$ with the associated $G(C)$.  Here we see that each $D(C)$ is associated with some particular crowding function, but some of the properties of these crowding functions are not as physically reasonable as those in Figures~\ref{fig:functions0}(a)--(b).  One attractive property of the crowding function in Figure~\ref{fig:functions0}(d) is that $G(1)=0$ for each case, and this is reasonable since we expect motility to cease when the lattice is packed to maximum density.  One less appealing feature of the crowding function in Figure~\ref{fig:functions0}(d) is that each case has $G(0)=0$, which means that isolated agents do not move, and clearly this is at odds with our intuition, and experimental evidence, that crowding reduces motility \citep{Lee1994inhibition,Tremel2009}. Despite this limitation, it is still insightful and interesting that we are able to take canonical choices of nonlinear diffusivity function $D(C)$, and to explore what choice of crowding function $G(C)$ leads to those nonlinear diffusivities. 

\section{Nonlinear diffusion influences population dynamics}
\label{sec_4}
In this section we quantify various population dynamics using both the discrete and associated continuum models. In all simulations, we consider an $L\times L$ domain where $L=100$, and we impose periodic boundary conditions along all boundaries. Agents are initially located in a central vertical strip of width $w$, which may represent a species along a river \citep{Lutscher2010persistence} or a population of cells in a scratch assay \citep{Jin2016reproducibility}. For the continuum model, since the initial distribution is independent of the vertical location and evolves with periodic boundary conditions, the population density remains independent of the vertical position for all $t > 0$ \citep{Simpson2009depth}. Therefore, Equation~\eqref{PDE_general_2D} simplifies to
\begin{equation}
    \label{PDE_general_1D}
    \frac{\partial C(x,t)}{\partial t}=\frac{\partial }{\partial x}\left(D(C(x,t))\frac{\partial C(x,t)}{\partial x}\right)+R(C(x,t)),
\end{equation}
where $C(x,t)$ represents the average column density of population \citep{simpson2010cell,Simpson2009depth}. We numerically solve Equation~\eqref{PDE_general_1D} and compute
\begin{equation}
    \label{totaldensity_1D}
    \mathcal{C}(t)=\frac{1}{L}\int_0^{L}C(x,t)\ \textrm{d}x,
\end{equation}
which is the total population density across the whole domain. We apply the method of lines to solve \eqref{PDE_general_1D} numerically and full details of the numerical method is given in the Supplementary Material. 

To quantify results from the discrete model we always consider performing $V$ identically prepared realisations, and use this data to calculate the average occupancy of site $\mathbf{s}$,
\begin{equation}
    \label{averageoccupancy}
    \bar{C}_{\mathbf{s}}=\frac{1}{V}\sum_{v=1}^{V}C^{(v)}_\mathbf{s}(t),
\end{equation}
where $C^{(v)}_\mathbf{s}(t)\in\{0,1\}$ is the occupancy of site $\mathbf{s}$ at time $t$ in the $v$th identically-prepared realisation. As the initial occupancy is independent of the vertical position, we can denote the average column density at time $t=n\tau$ as
\begin{equation}
\label{approximate_cxt}
    \langle C(x,t) \rangle=\frac{1}{VJ}\sum_{v=1}^{V}\sum_{j=1}^{J}C^{(v)}(i,j,n),
\end{equation}
which corresponds to $C(x,t)$ in the continuum model, where indexes $i$ and $j$, indicating the position of site $\mathbf{s}$, relate to position $(x,y)$. We also compute the total population density across the whole domain at time $t=n\tau$ as 
\begin{equation}
    \label{approximate_ct}
    \langle C(t) \rangle=\frac{1}{VIJ}\sum_{v=1}^{V}\sum_{j=1}^{J}\sum_{i=1}^{I}C^{(v)}(i,j,n),
\end{equation}
which corresponds to $\mathcal{C}(t)$ in the continuum model.

\begin{figure}
\centering
\includegraphics[width=\textwidth]{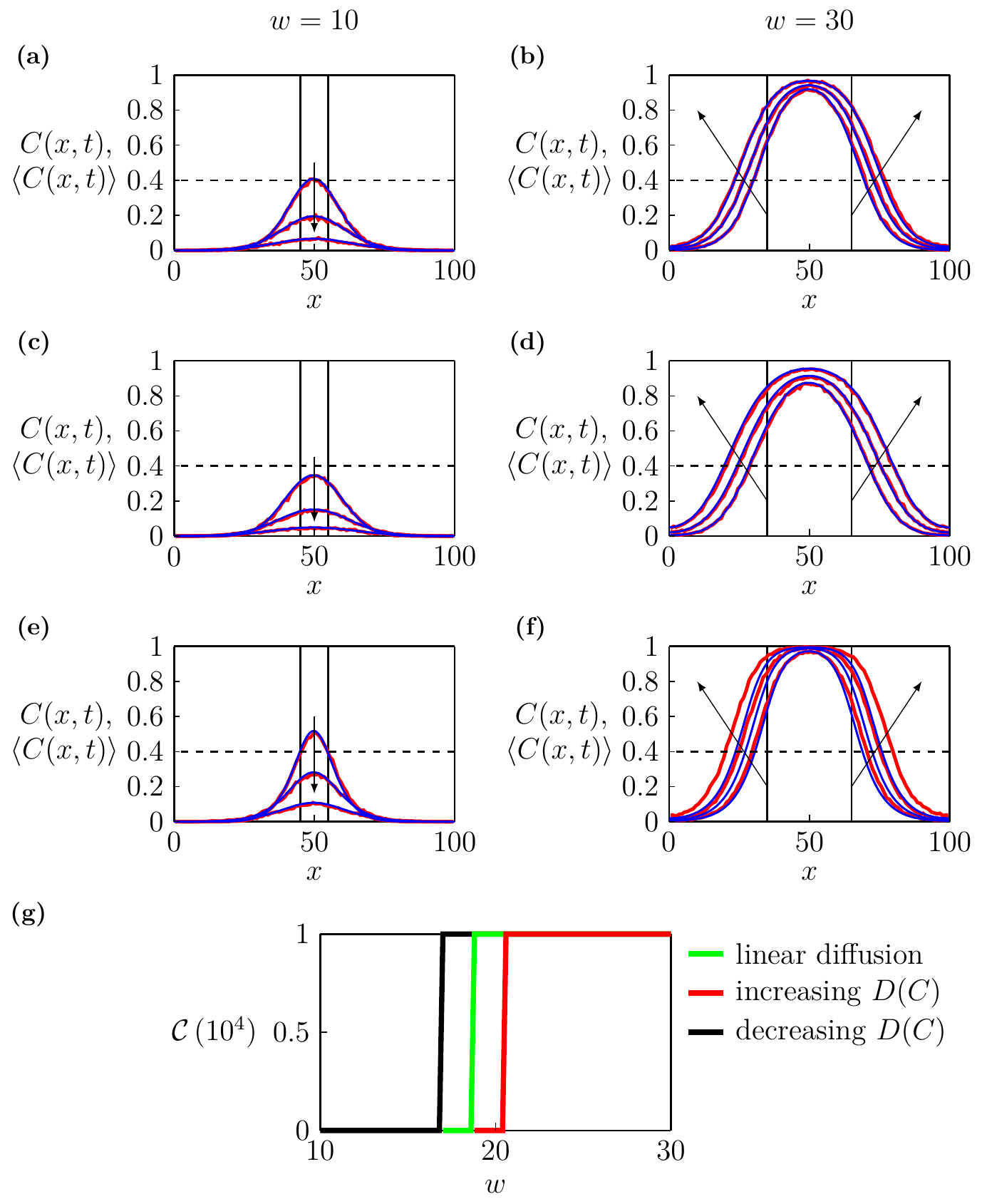}
\caption{\textbf{Comparisons of discrete and continuum results.} (a)--(f) $\langle C(x,t)\rangle$ (red) and $C(x,t)$ (blue). We use the linear diffusion given by \eqref{eq1} in (a)--(b), the increasing $D(C)$ given by \eqref{eq2} in (c)--(d), and the decreasing $D(C)$ given by \eqref{eq3} in (e)--(f). The solid black lines in (a), (c), (e) indicate that we consider an initial vertical strip with $w=10$.
Similarly, the solid black lines in (b), (d), (f) indicate that we consider an initial vertical strip with $w=30$. We show the column density at $t=200,400,600$ in (a), (c) and (e), at $t=800,1600,2400$ in (b) and (d), and at $t=1600,3200,4800$ in (f). The arrows in (a)--(f) show the direction of increasing time. (g) The total population density obtained from the continuum model, $\mathcal{C}(t)$, at $t=10^4$ with the linear diffusion (green), the increasing $D(C)$ (red) and the decreasing $D(C)$ (black).}
\label{fig:4} 
\end{figure}

Figures~\ref{fig:4}(a)-(b) show results from both the discrete and continuum models with $G(C)=1-C$ and $D(C) = D_0$, corresponding to linear diffusion. Discrete simulations are performed with $M=1$ and $P=6/1000$, leading to $D_0=1/4$ and $\lambda=6/1000$ in the continuum model. For the discrete model, we initially locate agents in the central vertical strip with width $w=10$, which means that the initial condition for \eqref{PDE_general_1D} is $C(x,0)=1$ for $x \in [45,55]$ and $C(x,0)=0$ elsewhere. Comparing $C(x,t)$ and $\langle C(x,t)\rangle$ shows that the match between the discrete and continuum results is excellent. Moreover, the density eventually reaches zero everywhere, which suggests that the population goes extinct. We then consider a larger width $w=30$, and compare $\left<C(x,t)\right>$ with $C(x,t)$ at $t=800,1600,2400$ in Figure~\ref{fig:4}(b). The solutions of \eqref{totaldensity_1D} also match well with the averaged data from discrete simulations. In this case, the column density eventually reaches the carrying capacity everywhere, which suggests that the population survives. 

Next, we consider the same initial conditions except that we use $G(C)=(1-C)(1+C/2)$ in the discrete model and the increasing $D(C)$, given by \eqref{eq2}, in the continuum model. Results in Figure~\ref{fig:4}(c)-(d) correspond to $w=10$ and $w=30$, respectively. Again, the continuum and discrete results match well. The population goes extinct with $w=10$, but survives with $w=30$. Comparing results in Figure~\ref{fig:4}(a)-(b) with those in Figure~\ref{fig:4}(c)-(d) shows that the evolution of $C(x,t)$ is different, and this difference is due to the role of nonlinear diffusion. Traditionally, if we were working with the continuum model alone, it would be difficult to provide a physical interpretation of these differences, but in our framework we can explain these differences through our simple crowding function, $G(C)$. We then use $G(C)=(1-C)(1-C/2)$ in the discrete model and the decreasing $D(C)$ given by \eqref{eq3} in the continuum model in Figures~\ref{fig:4}(e)--(f). In this case the continuum and discrete results reasonably match. Again, the population goes extinct with $w=10$, but survives with $w=30$.

Results in Figure~\ref{fig:4}(a)--(f) indicate the continuum limit of our discrete model provides an accurate approximation of the stochastic population dynamics for these three choices of crowding functions. In the Supplementary Material, we show that the discrete and continuum results also match well when we consider the power law diffusivity.  A natural question that arises when confronted with these results is the following: introducing a nonlinear diffusivity function changes the rate at which the population spreads across the domain, and we wish to understand how these differences affect the long-term survival or extinction of the bistable population. To begin to explore this question we now vary the initial width of the vertical strip $w\in[10,30]$, and show the total population density of the continuum model after a long period of time $t=10^4$, with the three $D(C)$ functions, given by \eqref{eq1}--\eqref{eq3} in Figure~\ref{fig:4}(g). Results in Figure~\ref{fig:4}(g) show that solutions of the continuum model with these three $D(C)$ functions lead to different critical values of $w$ that separate long-term survival from long-term extinction. This numerical exploration shows that the fate of bistable populations depends upon the choice of $D(C)$. We will now take advantage of our discrete-to-continuum framework to interpret how different choices of $D(C)$ either enhance or suppress population extinction.

\section{Interpretation of how \textit{D(C)} affects extinction}
\label{sec_5}
\begin{figure}[]
\centering
\includegraphics[width=0.6\textwidth]{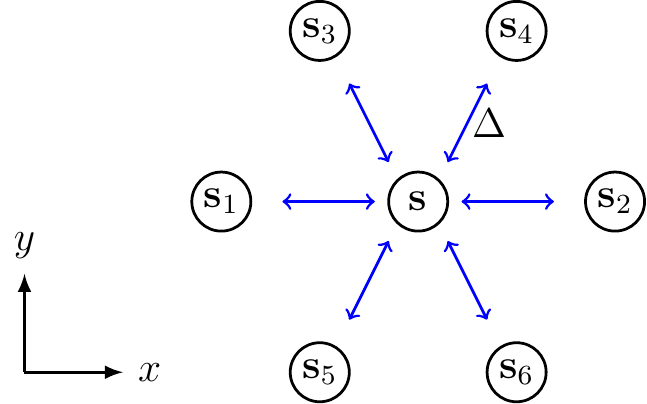}
\caption{\textbf{Schematic diagram showing the six neighbouring sites surrounding site $\mathbf{s}$.} The blue arrows indicate potential movement events that could change the occupancy of site $\mathbf{s}$. The spacing between site $\mathbf{s}$ and its neighbouring sites is $\Delta$.}
\label{fig:5} 
\end{figure}

To understand how different choices of $D(C)$ affect long-term survival or extinction, we now derive mathematical expressions for the average flux of agents in the discrete model in a general setting. We consider an agent at site $\mathbf{s}$, at location $(x,y)$, where the occupancy is $C(x,y,t)$. Assuming that the density is sufficiently smooth, the densities at the neighbouring sites can be obtained by expanding $C(x,y,t)$ in a truncated Taylor series,
\begin{equation}
    \nonumber
    \begin{aligned}
    &C_{1,2}=C\mp\Delta \frac{\partial C}{\partial x}+\mathcal{O}(\Delta^2),\\
    &C_{3,5}=C-\frac{\Delta }{2}\frac{\partial C}{\partial x}+\mathcal{O}(\Delta^2),\\ 
    &C_{4,6}=C+\frac{\Delta }{2}\frac{\partial C}{\partial x}+\mathcal{O}(\Delta^2),
    \end{aligned}
\end{equation}
where, for convenience, we denote $C(x,y,t)$ as $C$ and index the densities at neighbouring sites with subscripts as shown in Figure~\ref{fig:5}. 
The transition probability of an agent moving out of site $\mathbf{s}$ to one of its neighbouring sites $\mathbf{s}_i$, for $i=1,2,3, \ldots, 6$, is
\begin{equation}
\nonumber
    P^{-}_{i}=\frac{MCG(K)}{6(1-K)}(1-C_{i}).
\end{equation}
Similarly, the transition probability of an agent moving from site $\mathbf{s}_i$ into site $\mathbf{s}$ is
\begin{equation}
\nonumber
    P^{+}_{i}=\frac{MC_iG(K_i)}{6(1-K_i)}(1-C).
\end{equation}
Therefore, combining these expressions for the transition probabilities with the geometry of the lattice in Figure~\ref{fig:4}, allows us to write down an expression for the horizontal component of the net flux of agents at site $\mathbf{s}$,
\begin{equation}
    \label{interface2}
    \begin{aligned}
    J_x=&\frac{\Delta}{2\tau}\left[\left(2 P^{-}_{2}+P^{-}_{4}+P^{-}_6\right)+\left(2 P^{+}_{1}+P^{+}_{3}+P^{+}_5\right)\right]\\
    &-\frac{\Delta}{2\tau}\left[\left(2 P^{+}_{2}+P^{+}_{4}+P^{+}_6\right)+\left( 2P^{-}_{1}+P^{-}_{3}+P^{-}_5\right)\right].
    \end{aligned}
\end{equation}
Substituting the expressions for the transition probabilities into \eqref{interface2} and then expanding the resulting terms in truncated Taylor series about site $\mathbf{s}$ gives
\begin{equation}
    \label{interface3}
    J_x=-\frac{M\Delta^2 }{4\tau}\left[C\frac{\textls{d}G(C)}{\textls{d}C}+\frac{1+C}{1-C}G(C)\right]\frac{\partial C}{\partial x}+\mathcal{O}(\Delta^3),
\end{equation}
We note that \eqref{interface3} can be written as
\begin{equation}
    \label{interface3_written}
    \begin{aligned}
    J_x=-D(C)\frac{\partial C}{\partial x},
    \end{aligned}
\end{equation} 
where $D(C)$ is the same as \eqref{relation}. Following a similar conservation argument, the flux of agents in the vertical direction can be written as
\begin{equation}
    \label{interface3_vertical}
    J_y
    =-D(C)\frac{\partial C}{\partial y}.
\end{equation}
For all simulations in this work our initial condition is independent of vertical position so we have $J_y=0$ throughout. It is useful to compare the nonlinear diffusive flux term with the classical linear diffusion flux. Thus, we re-write \eqref{interface3_written} as 
\begin{equation}
    \label{eqKc_3} 
    J_x=-D_0\left(1+H(C)\right)\frac{\partial C}{\partial x},
\end{equation}
where $H(C)$ can be regarded as a correction that is associated with the effects of nonlinear diffusion. For example, setting $H(C)=0$ means that out nonlinear diffusion term simplifies to the classical linear diffusion term, whereas setting $H(C) > 0$ means that the nonlinear diffusion term is larger than the associated linear diffusion term.

\begin{landscape}
\begin{figure}
\centering
\includegraphics[width=1.2\textwidth]{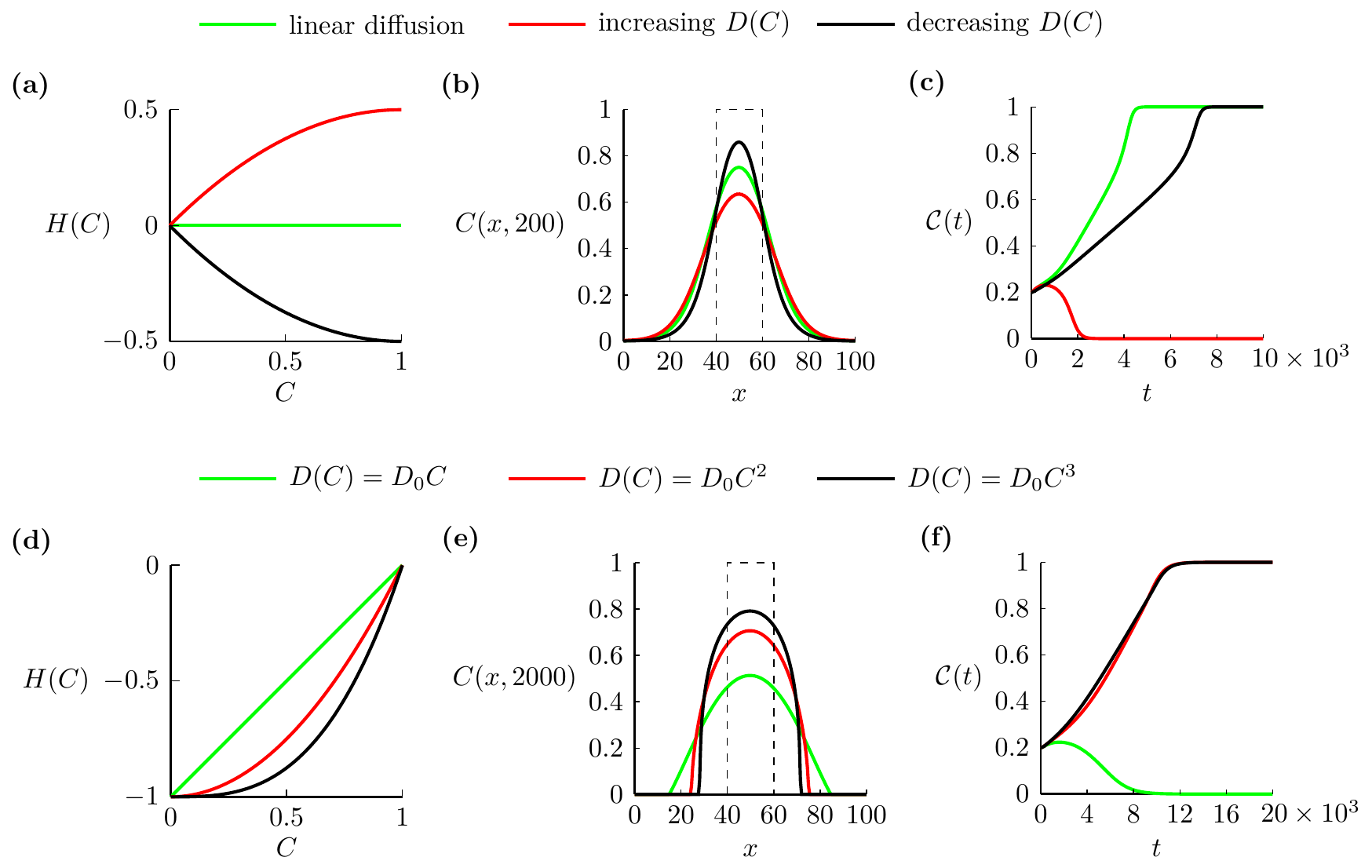}
\caption{\textbf{Population dynamics with different $D(C)$.} (a) $H(C)$ associated with the linear diffusion given by \eqref{eq1} (green), increasing $D(C)$ given by \eqref{eq2} (red), and decreasing $D(C)$ given by \eqref{eq3} (black). (b) Solutions of Equation \eqref{PDE_general_1D} at $t=200$ with these three $D(C)$ functions. The dashed line is the initial distribution where $C(x,0)=1$ for $x\in[40,60]$ and $C(x,0)=0$ elsewhere. (c) The evolution of $\mathcal{C}(t)$ with these three $D(C)$ functions. We set $M=1$ and $P=6/1000$ leading to $D_0=1/4$ and $\lambda=6/1000$ in (b)--(c). (d) $H(C)$ associated with $D(C)=D_0C$ (green), $D(C)=D_0C^2$ (red) and $D(C)=D_0C^3$ (black). (e) Solutions of Equation \eqref{PDE_general_1D} at $t=2000$ with these three $D(C)$ functions. The dashed line is the initial distribution where $C(x,0)=1$ for $x\in[40,60]$ and $C(x,0)=0$ elsewhere. (f) The evolution of $\mathcal{C}(t)$ with these three $D(C)$ functions. We set $M=1$ and $P=1/1000$ leading to $D_0=1/4$ and $\lambda=1/1000$ in (e)--(f).}
\label{fig:columnl}
\end{figure}
\end{landscape}

To explore how different choices of $D(C)$ affect the long-term fate of bistable populations, we repeat the kinds of simulations we considered in Figure~\ref{fig:4}, and summarise the results in Figure~\ref{fig:columnl} where we consider the linear diffusion, increasing $D(C)$ and decreasing 
$D(C)$ functions given by \eqref{eq1}--\eqref{eq3}, respectively.
For each diffusivity function we plot $H(C)$ in Figure~\ref{fig:columnl}(a). The increasing $D(C)$ leads to $H(C)\ge0$ showing that the nonlinear flux is greater than the flux associated with the linear diffusion model for $C \in (0,1]$. In contrast, the decreasing $D(C)$ leads to $H(C)\le0$ showing that the nonlinear flux is less than the flux associated with the linear diffusion model for $C \in (0,1]$. We consider an initial condition with width $w=20$ together with $M=1$ and $P=6/1000$, which corresponds to $D_0=1/4$ and $\lambda=6/1000$ in the continuum model. Profiles in Figure~\ref{fig:columnl}(b) show the solution of the continuum model at $t=200$, where we see that the density profile associated with the increasing $D(C)$ spreads further than the linear diffusion model.  Similarly, the profile associated with the decreasing $D(C)$ spreads less than the linear diffusion model. These differences in spatial spreading mean that the maximum density for the increasing $D(C)$ model is less than the maximum density for the linear diffusion model, which encourages extinction since the bistable source term becomes negative across a larger area of the domain.  This is consistent with the results in Figure~\ref{fig:columnl}(c) showing the long-term evolution of $\mathcal{C}(t)$, where we see that the population with increasing $D(C)$ goes extinct whereas the populations with linear diffusion and decreasing $D(C)$ lead to long-term survival.

Next, we investigate the population dynamics with the nonlinear diffusivity functions $D(C) = D_0 C^m$ for $m=1,2$ and $3$, with the associated $H(C)$ functions in Figure~\ref{fig:columnl}(d). With the same initial conditions and discrete parameters in Figure 6, we show various solutions of \eqref{PDE_general_1D} at time $t=2000$ in Figure~\ref{fig:columnl}(e), and $\mathcal{C}(t)$ in Figure~\ref{fig:columnl}(f).
The results in Figure~\ref{fig:columnl}(d)--(f) for this class of power law $D(C)$ are consistent with the results in Figure~\ref{fig:columnl}(a)--(c).  Results for $m=1,2$ and $3$ lead to a reduction in flux relative to linear diffusion, but setting $m=1$ still leads to sufficient spreading that a larger proportion of the domain has $C < A$ leading to extinction, whereas setting $m > 1$ reduces the spreading so that a smaller proportion of the domain has $C < A$ leading to survival.

Results in Figure~\ref{fig:columnl} show that the change of flux influences the speed at which the population spreads in space. Due to the change of spreading speed, the initial width of the vertical strip needed for a population to survive changes as well. This suggests that the nonlinear diffusivity function affects the fate of bistable populations through influencing the flux of populations. However, in Figure~\ref{fig:columnl}, we fix the ratio of growth and movement, $P/M$, while $P/M$ also influences the fate of bistable populations. Next, we are going to vary $P/M$ and study the influence of nonlinear diffusion on the fate of a broader range of populations.

\begin{figure}[]
\centering
\includegraphics[width=\textwidth]{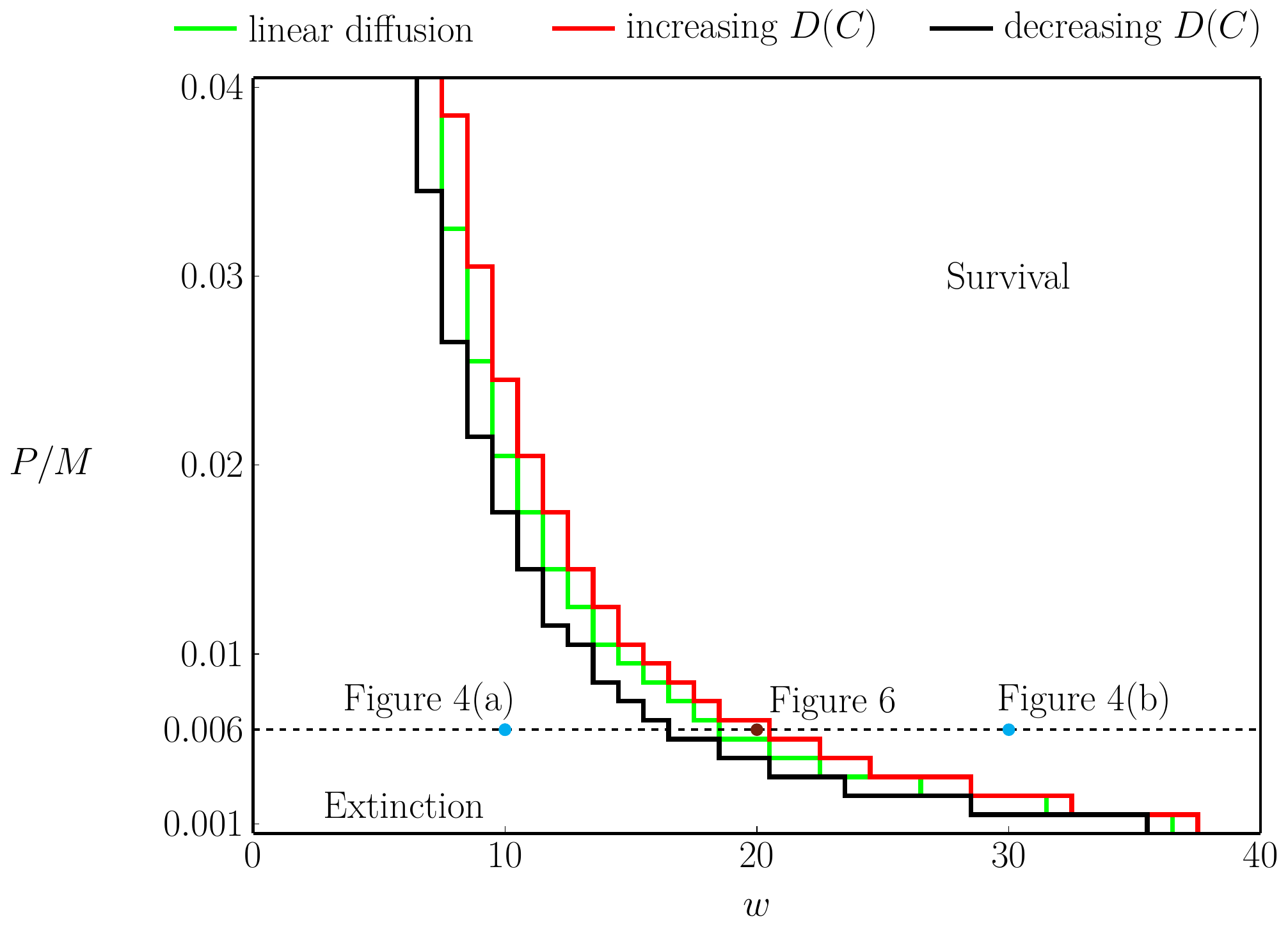}
\caption{\textbf{Phase diagram for the survival and extinction of populations.} The phase space $(w,P/M)$ is uniformly discretised into a rectangular mesh with $41\times40$ nodes, where $w\in[0,40]$ and $P/M\in[1/1000,4/100]$. Three curves are the thresholds for the survival and extinction of populations in the continuum model with the linear diffusion given by~\eqref{eq1} (green), increasing $D(C)$ given by~\eqref{eq2} (red), and decreasing $D(C)$ given by~\eqref{eq3} (black). Two cyan dots indicate the parameters $P/M=6/1000$ and $w=10$ or $w=30$ considered in Figures~\ref{fig:4}. The brown dot indicates the parameters $P/M=6/1000$ and $w=20$ considered in Figure~\ref{fig:columnl}.}
\label{fig:PhaseDiagram_Num_1D} 
\end{figure}

Our results so far show that the long-term survival of the population involves a complicated relationship between the width of the initial condition $w$, the time scale of migration $M$, the time scale of growth $P$, as well as the particular nonlinear diffusivity function $D(C)$. We now systematically explore this relationship by taking the $(w, P/M)$ phase space and discretising it uniformly for $w \in [0,40]$ and $P/M \in [1/1000, 40,1000]$, as shown in Figure~\ref{fig:PhaseDiagram_Num_1D}. We vary $P$ and fix $M=1$ leading to $\lambda=P$ and $D_0=1/4$ in the continuum model. As we are interested in the long-term outcomes of bistable populations, we solve \eqref{PDE_general_1D} and calculate $\mathcal{C}(T)$ for a sufficiently long period of time $T$, so that the outcome is either $\mathcal{C}(T)=1$ or $\mathcal{C}(T)=0$. With these simulation outcomes we identify the boundaries that separate survival and extinction on the phase diagram.
Overall, results in Figure~\ref{fig:PhaseDiagram_Num_1D} show that large $w$ encourages survival, and that for each value of $w$ there is a threshold value of $P/M$ that determines the eventual survival or extinction of the survival. In the Supplementary, we show that the boundaries generated from the discrete model are consistent with the boundaries identified using the continuum model. The main result in Figure~\ref{fig:PhaseDiagram_Num_1D} is that the curves that delineate the survival/extinction boundary depend upon the choice of $D(C)$, and we plot three curves for the linear diffusion model, the increasing $D(C)$ given by \eqref{eq2} and the decreasing $D(C)$ given by \eqref{eq3}.  The horizontal dashed line at $P/M = 0.006$ highlights results shown previously in Figure~\ref{fig:4} and Figure~\ref{fig:columnl}, but this phase diagram summarises the long-term survival/extinction patterns for a much wider choice of parameters than we explored in these previous cases. 

\section{Conclusion and future work}
In this work we consider the question of population survival or extinction, with a focus on understanding how various migration mechanisms either encourage or suppress extinction.  In the population biology modelling literature, the most common way to study population dynamics with spatial effects is to use a reaction-diffusion model with a linear diffusion term to represent migration and a bistable source term to represent birth-death processes.  While most studies employ a linear diffusion mechanism for simplicity, there are many cases where linear diffusion is inadequate.  For example, mathematical models based on a linear diffusion mechanism do not predict a well-defined front that is often observed experimentally or in the field.  This limitation of linear diffusion is typically overcome by generalising the constant diffusivity, $D$, to a nonlinear diffusivity function $D(C)$.  One of the main challenges of working with a nonlinear diffusion framework is the important question of how to choose the functional form of $D(C)$, and there are conflicting results in the literature.  For example, in the cell migration modelling literature some studies have found that using a power law diffusivity function $D(C) = D_0 C^m$, with $m \ge 1$ can lead to a good match to experimental data \citep{Sherratt1990,Sengers2007,Jin2016reproducibility}. One of the features of these models it that this nonlinear diffusivity is an increasing function of density.  Curiously, other researchers working in precisely the same field have suggested that a decreasing nonlinear diffusivity function is appropriate, $D(C) = 1/(\alpha+C)$, with $\alpha >0$ \citep{Cai2007}.  This highlights the fact that choosing an appropriate nonlinear diffusivity function is not always straightforward.

\begin{figure}[]
\centering
\includegraphics[width=0.8\textwidth]{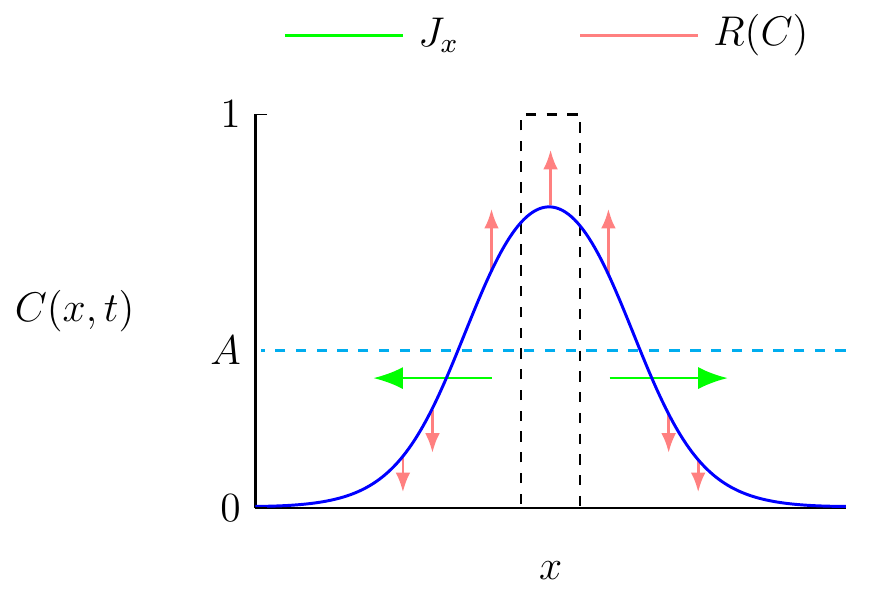}
\caption{\textbf{Schematic profile of the spreading population (blue) superimposed on the Allee threshold $C=A$ (dashed cyan) and the initial density distribution $C(x,0)$ (dashed black)}.  Spatial spreading of the population is controlled by the diffusive flux, $J_x$, that is proportional to the nonlinear diffusivity function, $D(C)$, and the direction of the flux is indicated (green arrows).  This flux affects: (i) the proportion of the domain where $C<A$ giving rise to a negative source term (downward red arrows), and (ii) the proportion of the domain where $C > A$, giving rise to a positive source term (upward red arrows).}
\label{fig:8} 
\end{figure}

In addition to understanding how to choose an appropriate nonlinear diffusivity function, a related challenge is to understand how different forms of $D(C)$ affect the long-term survival or extinction of bistable populations.  While it has been established that different choices of $D(C)$ impacts the long-term survival of populations \citep{lee2006bifurcation}, an intuitive understanding of why different choices of $D(C)$ encourage or suppress extinction has been lacking.  In this work we address this question by working with a very simple discrete modelling framework on a two-dimensional hexagonal lattice, where migration and birth/death events are controlled through relatively simple, easy-to-interpret crowding functions. In particular we work with a migration crowding function $G(C)$, which provides a very simple measure of how the ability of an individual agent to move is reduced as a function of density, $C$.  Our discrete-to-continuum averaging arguments provides a mathematical relationship which allows us to either: (i) specify $G(C)$ and determine the associated nonlinear diffusivity function $D(C)$; or, (ii) specify $D(C)$ and determine the associated crowding function $G(C)$. This new relationship allows us to explore how the averaged population-level flux of agents varies relative to the classical linear diffusion model for a particular crowding function, $G(C)$.  We find that choices of $G(C)$ that increase the flux encourage population extinction relative to the classical linear diffusion model, whereas choices of $G(C)$ that decrease the flux suppresses population extinction.  These results are summarised in the conceptual diagram in Figure~\ref{fig:8} showing that, for the initial conditions considered, increasing the flux of agents tends to reduce the density across the domain as the population spreads, meaning that a greater proportion of the domain has $C < A$, where the bistable source term acts to reduce the population and encourage extinction.  

There are many ways that our work can be extended.  For example, all simulation results presented here consider a very simple one-dimensional vertical strip initial condition. These results can be generalised to other initial shapes, such as circular, square or more complicated initial populations, and the mathematical and computational tools presented in this work can be applied directly to this generalisation. We show that nonlinear diffusion plays the same role as linear diffusion on the fate of populations when we consider the simple well-mixed initial distribution in the Supplementary Material. There are also many ways that the current modelling framework can be extended.  For example, the discrete model can be extended to consider multiple interacting subpopulations, and the same discrete-to-continuum averaging approach could be used to construct a continuum limit model, which would take the form of a system of coupled partial differential equation models \citep{shigesada1979,Painter2003,Hughes2010attracted,Ricardo2020}. We leave this extension for future consideration.  

\hfill

\noindent
\textit{Acknowledgements:} This work is supported by the Australian Research Council (DP200100177, DP190102545). We appreciate advice from Professor Petrus van Heijster and Dr Stuart Johnston.

\appendix 

\renewcommand{\thealgocf}{S\arabic{algocf}}

\newpage
\section{Algorithm for discrete simulations}
We consider an $L\times L$ domain with $L=100$. Each lattice site is indexed by $(i,j)$,  and has a unique Cartesian coordinate, 
\begin{equation}
    \label{xyijrelation}
    (x,y)=\left\{\begin{aligned}
    &\left(i\Delta,j\dfrac{\Delta\sqrt{3}}{2}\right),\quad && \text{if $j$ is even,}\\
    &\left(\left(i+\dfrac{1}{2}\right)\Delta,j\dfrac{\Delta\sqrt{3}}{2}\right),\quad && \text{if $j$ is odd.}
    \end{aligned}\right.
\end{equation}
To approximate the size of the domain in discrete simulations, we create a two-dimensional hexagonal lattice with  $100\times116$ uniformly distributed nodes. The pseudo-code for a single realisation of the stochastic model is given in Algorithm \ref{algorithm1}.

\setstretch{0.9}
\begin{tcolorbox}
\begin{algorithm}[H]
\SetAlgoLined
 Create a two-dimensional $I\times J$ hexagonal lattice; Distribute agents with vertical strip initial conditions; The total number of lattice site is $IJ$\;
 Set $t=0$; Calculate total agents $Q(t)$\;
 \While{\normalfont $t<t_{\text{end}}$ and $Q(t)>0$ and $Q(t)\le IJ$  }{
 $t=t+\tau$\;
 $Q(t)=Q(t-\tau)$\;
 $B_1=0$; $B_2=0$\;
 Draw two random variables: $\beta_1\sim \textit{U}[0,1]$, $\beta_2\sim \textit{U}[0,1]$\;
 \While{$B_1<Q(t)$ }{
 $B_1=B_1+1$\;
 Randomly choose an agent $\mathbf{s}$\;
  \uIf{$\beta_1<M$}{
   Calculate $\newbar{K}_{\mathbf{s}}^{(\textls{m})}$ and $G(\newbar{K}_{\mathbf{s}}^{(\textls{m})})$\;
   Draw a random variable: $\gamma_1\sim \textit{U}[0,1]$\;
    \uIf{\normalfont $\gamma_1<G(\newbar{K}_{\mathbf{s}}^{(\text{m})})$}{
       Randomly choose a vacant site in $\mathcal{N}_{1}(\mathbf{s})$ and move agent to chosen site
       }
    \Else{
   Nothing happens\;
  }
  }
  \Else{
   Nothing happens\;
  }
 }
  \While{$B_2<Q(t)$ }{
 $B_2=B_2+1$\;
 Randomly choose an agent $\mathbf{s}$\;
  \uIf{$\beta_2<P$}{
   Calculate ${K}_{\mathbf{s}}^{(\textls{g})}$ and $F(\newbar{K}_{\mathbf{s}}^{(\textls{g})})$\;
   Draw a random variable: $\gamma_2\sim \textit{U}[0,1]$\;
   \uIf{\normalfont $F(\newbar{K}_{\mathbf{s}}^{(\text{g})})>0$}{
       \uIf{\normalfont $\gamma_2<F(\newbar{K}_{\mathbf{s}}^{(\text{g})})$}{
       Randomly choose a vacant site in $\mathcal{N}_{4}(\mathbf{s})$ and place a new agent on chosen site\;
       $Q(t)=Q(t)+1$
       }
   }
   \uElseIf{\normalfont $F(\newbar{K}_{\mathbf{s}}^{(\text{g})})<0$}{
    \uIf{\normalfont $\gamma_2<-F(\newbar{K}_{\mathbf{s}}^{(\text{g})})$}{
       Remove agent\;
       $Q(t)=Q(t)-1$\;
       }
  }
    \Else{
   Nothing happens\;
  }
  }
  \Else{
   Nothing happens\;
  }
 }
 }
\caption{Pseudo-code for a single realisation of the stochastic model}
\label{algorithm1}
\end{algorithm}
\end{tcolorbox}

\newpage
\setstretch{1.5}
\section{Derivation of the continuum limit}
Considering the spatial template with $r=1$ for the movement mechanism and $r\ge1$ for the growth mechanism of agents, the expected change in occupancy of site $\mathbf{s}$ during the time interval from $t$ to $t+\tau$ is given as
\begin{equation}
\label{delta1}
     \begin{aligned}
     \delta(\newbar{C}_{\mathbf{s}})=&\frac{M}{\lvert\mathcal{N}_{1}\rvert}(1-\newbar{C}_{\mathbf{s}})\sum_{\mathbf{s}'\in \mathcal{N}_{1}\{\mathbf{s}\}}\newbar{C}_{\mathbf{s}'}\frac{G(\newbar{K}_{\mathbf{s}'}^{(\textls{m})})}{1-\newbar{K}_{\mathbf{s}'}^{(\textls{m})}}-M\newbar{C}_{\mathbf{s}}G(\newbar{K}_{\mathbf{s}}^{(\textls{m})})\\
          &+\frac{P}{\lvert\mathcal{N}_{r}\rvert}(1-\newbar{C}_{\mathbf{s}})\sum_{\mathbf{s}'\in \mathcal{N}_{r}\{\mathbf{s}\}}\mathbbm{H}(F(\newbar{K}_{\mathbf{s}'}^{(\textls{g})}))\newbar{C}_{\mathbf{s}'}\frac{F(\newbar{K}_{\mathbf{s}'}^{(\textls{g})})}{1-\newbar{K}_{\mathbf{s}'}^{(\textls{g})}}-(1-\mathbbm{H}(F(\newbar{K}_{\mathbf{s}}^{(\textls{g})}))P\newbar{C}_{\mathbf{s}}F(\newbar{K}_{\mathbf{s}}^{(\textls{g})}).
     \end{aligned}
\end{equation}
As we know that the continuum limit of the last two terms in Equation \eqref{delta1} leads to a source term $\lambda C F(C)$ with $r\ge1$ \citep{Jin2016Sto}, we focus on the movement mechanism, that is, the first two terms on the right hand side of Equation~\eqref{delta1}. For convenience, we will omit the overlines on notations in the following content. 

It is useful to first write the general form of the Taylor series relating the occupancy of sites $(x+a,y+b)$,
\begin{equation}
    \label{general}
    C_{x+a,y+b}=C_{x,y}+\frac{(a\Delta)^1}{1!}\frac{\partial C_{x,y}}{\partial x}+\frac{(b\Delta)^1}{1!}\frac{\partial C_{x,y}}{\partial y}+\frac{(a\Delta)^2}{2!}\frac{\partial C^2_{x,y}}{\partial x^2}+\frac{2ab\Delta^2}{2!}\frac{\partial C^2_{x,y}}{\partial x\partial y}+\frac{(b\Delta)^2}{2!}\frac{\partial C^2_{x,y}}{\partial y^2}+\mathcal{O}(\Delta^3).
\end{equation}
We represent the six nearest neighbouring sites of site~$\mathbf{s}$ located at $(x,y)$ as site $\mathbf{s}_1$ with $(x-\Delta,y)$; site $\mathbf{s}_2$ with $(x+\Delta,y)$; site $\mathbf{s}_3$ with $(x-\Delta/2,y+\Delta\sqrt{3}/2)$; site $\mathbf{s}_4$ with $(x+\Delta/2,y+\Delta\sqrt{3}/2)$; site $\mathbf{s}_5$ with $(x-\Delta/2,y-\Delta\sqrt{3}/2)$ and site $\mathbf{s}_6$ with $(x+\Delta/2,y-\Delta\sqrt{3}/2)$. That is, $\mathcal{N}_1=\{\mathbf{s}_1,\mathbf{s}_2,\mathbf{s}_3,\mathbf{s}_4,\mathbf{s}_5,\mathbf{s}_6\}$. The truncated Taylor series of these sites are
\begin{align}
    &C_{\mathbf{s}_1}=C_{\mathbf{s}}-\frac{\partial C_{\mathbf{s}}}{\partial x}\Delta+\frac{\partial^2 C_{\mathbf{s}}}{\partial x^2}\frac{\Delta^2}{2}+\mathcal{O}(\Delta^3),\label{Tayler1}\\
    &C_{\mathbf{s}_2}=C_{\mathbf{s}}+\frac{\partial C_{\mathbf{s}}}{\partial x}\Delta+\frac{\partial^2 C_{\mathbf{s}}}{\partial x^2}\frac{\Delta^2}{2}+\mathcal{O}(\Delta^3),\label{Taylor2}\\
    &C_{\mathbf{s}_3}=C_{\mathbf{s}}-\frac{\partial C_{\mathbf{s}}}{\partial x}\frac{\Delta}{2}+\frac{\partial C_{\mathbf{s}}}{\partial y}\frac{\sqrt{3}\Delta}{2}+\left[\frac{1}{4}\frac{\partial^2 C_{\mathbf{s}}}{\partial x^2}+\frac{3}{4}\frac{\partial^2 C_{\mathbf{s}}}{\partial y^2}-\frac{\sqrt{3}}{2}\frac{\partial^2 C_{\mathbf{s}}}{\partial x\partial y}\right]\frac{\Delta^2}{2}+\mathcal{O}(\Delta^3),\label{Tayler3}\\
    &C_{\mathbf{s}_4}=C_{\mathbf{s}}+\frac{\partial C_{\mathbf{s}}}{\partial x}\frac{\Delta}{2}+\frac{\partial C_{\mathbf{s}}}{\partial y}\frac{\sqrt{3}\Delta}{2}+\left[\frac{1}{4}\frac{\partial^2 C_{\mathbf{s}}}{\partial x^2}+\frac{3}{4}\frac{\partial^2 C_{\mathbf{s}}}{\partial y^2}+\frac{\sqrt{3}}{2}\frac{\partial^2 C_{\mathbf{s}}}{\partial x\partial y}\right]\frac{\Delta^2}{2}+\mathcal{O}(\Delta^3),\label{Tayler4}\\
    &C_{\mathbf{s}_5}=C_{\mathbf{s}}-\frac{\partial C_{\mathbf{s}}}{\partial x}\frac{\Delta}{2}-\frac{\partial C_{\mathbf{s}}}{\partial y}\frac{\sqrt{3}\Delta}{2}+\left[\frac{1}{4}\frac{\partial^2 C_{\mathbf{s}}}{\partial x^2}+\frac{3}{4}\frac{\partial^2 C_{\mathbf{s}}}{\partial y^2}+\frac{\sqrt{3}}{2}\frac{\partial^2 C_{\mathbf{s}}}{\partial x\partial y}\right]\frac{\Delta^2}{2}+\mathcal{O}(\Delta^3),\label{Tayler5}\\
    &C_{\mathbf{s}_6}=C_{\mathbf{s}}+\frac{\partial C_{\mathbf{s}}}{\partial x}\frac{\Delta}{2}-\frac{\partial C_{\mathbf{s}}}{\partial y}\frac{\sqrt{3}\Delta}{2}+\left[\frac{1}{4}\frac{\partial^2 C_{\mathbf{s}}}{\partial x^2}+\frac{3}{4}\frac{\partial^2 C_{\mathbf{s}}}{\partial y^2}-\frac{\sqrt{3}}{2}\frac{\partial^2 C_{\mathbf{s}}}{\partial x\partial y}\right]\frac{\Delta^2}{2}+\mathcal{O}(\Delta^3).\label{Tayler6}
\end{align}
The local density of $\mathbf{s}$ is obtained by summing the Taylor series of sites in $\mathcal{N}_1\{\mathbf{s}\}$, that is,
\begin{equation}
\label{averages_c}
\begin{aligned}
    K_{\mathbf{s}}^{(\textls{m})}&=\frac{1}{6}\sum_{\mathbf{s}''\in\mathcal{N}_1\{\mathbf{s}\}}C_{\mathbf{s}''}\\
    &=C_{\mathbf{s}}+\left(\frac{\partial^2C_{\mathbf{s}}}{\partial x^2}+\frac{\partial^2C_{\mathbf{s}}}{\partial y^2}\right)\frac{\Delta^2}{4}+\mathcal{O}(\Delta^3).
\end{aligned}
\end{equation}
Similarly, the local density of $\mathbf{s_1}$ is obtained by summing the Taylor series of sites in $\mathcal{N}_1\{\mathbf{s}_1\}$, that is,
\begin{equation}
\label{averages'}
\begin{aligned}
    K_{\mathbf{s}_1}^{(\textls{m})}&=\frac{1}{6}\sum_{\mathbf{s}''\in\mathcal{N}_1\{\mathbf{s}_1\}}C_{\mathbf{s}''}\\
    &=C_{\mathbf{s}_1}+\left(\frac{\partial^2C_{\mathbf{s}_1}}{\partial x^2}+\frac{\partial^2C_{\mathbf{s}_1}}{\partial y^2}\right)\frac{\Delta^2}{4}+\mathcal{O}(\Delta^3),\\
    &=C_{\mathbf{s}}-\frac{\partial C_{\mathbf{s}}}{\partial x}\Delta+\frac{\partial^2C_{\mathbf{s}}}{\partial x^2}\frac{\Delta^2}{2}+\left(\frac{\partial^2C_{\mathbf{s}}}{\partial x^2}+\frac{\partial^2C_{\mathbf{s}}}{\partial y^2}\right)\frac{\Delta^2}{4}+\mathcal{O}(\Delta^3).
\end{aligned}
\end{equation}
For simplification we rewrite Equation \eqref{averages'} as $K_{\mathbf{s}_1}^{(\textls{m})}=C_\mathbf{s}+\widetilde{C}_{\mathbf{s}_1}$, where $\widetilde{C}_{\mathbf{s}_1}\sim\mathcal{O}(\Delta)$. Subsequently, the movement crowding function at $\mathbf{s}_1$ can be expanded as
\begin{equation}
    \label{crowding1}
    \begin{aligned}
        G\left(K_{\mathbf{s}_1}^{(\textls{m})}\right)&=G\left(C_\mathbf{s}+\widetilde{C}_{\mathbf{s}_1}\right),\\
        &=G\left(C_\mathbf{s}\right)+\frac{\textrm{d}G\left(C_\mathbf{s}\right)}{\textrm{d}C}\widetilde{C}_{\mathbf{s}_1}+\frac{\textrm{d}^2G\left(C_\mathbf{s}\right)}{\textrm{d}C^2}\frac{{\widetilde{C}_{\mathbf{s}_1}}^2}{2}.
    \end{aligned}
\end{equation}
The expansions of $G(K_{\mathbf{s}_2}^{(\textls{m})})$, $G(K_{\mathbf{s}_3}^{(\textls{m})})$,...,$G(K_{\mathbf{s}_6}^{(\textls{m})})$ have similar forms to \eqref{crowding1}. We then go back to the first term on the right hand side of \eqref{delta1}, which gives
\begin{equation}
\label{first_1}
        \frac{M}{6}(1-C_\mathbf{s})\sum_{\mathbf{s}'\in \mathcal{N}_1\{\mathbf{s}\}}C_{\mathbf{s}'}\frac{G(K_{\mathbf{s}'}^{(\textls{m})})}{1-K_{\mathbf{s}'}^{(\textls{m})}}.
\end{equation}
For convenience we further drop the $\mathbf{s}$ notation so that $C_\mathbf{s}$ becomes $C$ and $C_{\mathbf{s}_1}$ becomes $C_{1}$. Subsequently, \eqref{first_1} becomes
\begin{equation}
\label{first_2}
        \frac{M}{6}(1-C)\sum_{i=1}^6C_i\frac{G(K_{\mathbf{s}_i}^{(\textls{m})})}{1-K_{\mathbf{s}_i}^{(\textls{m})}}.
\end{equation}
Moreover, we will use two notations
\begin{equation}
    \mathcal{A}=\left(\frac{\partial^2C_{\mathbf{s}}}{\partial x^2}+\frac{\partial^2C_{\mathbf{s}}}{\partial y^2}\right)\frac{\Delta^2}{4}, \quad 
    \mathcal{B}=\left(\left(\frac{\partial C_{\mathbf{s}}}{\partial x}\right)^2+\left(\frac{\partial C_{\mathbf{s}}}{\partial y}\right)^2\right)\frac{\Delta^2}{4},
\end{equation}
in the following content. Expanding the term related to site $\mathbf{s}_1$ in \eqref{first_2} gives
\begin{equation}
    \nonumber
    \begin{aligned}
    &\dfrac{M}{6}(1-C)\left(C+\widetilde{C}_1-\mathcal{A}\right)\frac{\left(G(C)+G'(C)\widetilde{C}_1+G''(C)\dfrac{\widetilde{C}^2_1}{2}\right)}{1-\left(C+\widetilde{C}_1\right)}\\
    =&\dfrac{M}{6}(1-C)\left(C+\widetilde{C}_1-\mathcal{A}\right){\left(G(C)+G'(C)\widetilde{C}_1+G''(C)\dfrac{\widetilde{C}_1^2}{2}\right)}\left(\frac{1}{1-C}+\frac{{\widetilde{C}_1}}{(1-C)^2}+\frac{{\widetilde{C}_1}^2}{(1-C)^3}\right)+\mathcal{O}(\Delta^3)\\
    =&\dfrac{M}{6}\left(C+\widetilde{C}_1-\mathcal{A}\right){\left(G(C)+G'(C)\widetilde{C}_1+G''(C)\dfrac{\widetilde{C}_1^2}{2}\right)}\left(1+\frac{{\widetilde{C}_1}}{1-C}+\frac{{\widetilde{C}_1}^2}{(1-C)^2}\right)+\mathcal{O}(\Delta^3)\\
    =&\dfrac{M}{6}\left[CG(C)+\left(CG'(C)+\frac{G(C)}{1-C}\right)\widetilde{C}_1+\left(\frac{G(C)}{(1-C)^2}+\dfrac{G'(C)}{1-C}+\frac{CG''(C)}{2}\right){\widetilde{C}_1^2}-{G(C)}\mathcal{A}\right]+\mathcal{O}(\Delta^3).
    \end{aligned}
\end{equation}
Here, the prime denotes the ordinary differentiation with respect to $C$. The terms related to other sites can be obtained in a similar way. Therefore, expanding all terms in \eqref{first_2} and neglecting terms of order $\mathcal{O}(\Delta^3)$ gives
\begin{equation}
    \label{first_3}
    \dfrac{M}{6}\left[6CG(C)+\left(CG'(C)+\frac{G(C)}{1-C}\right)\sum_{k=1}^{6}\widetilde{C}_k+\left(\frac{G(C)}{(1-C)^2}+\dfrac{G'(C)}{1-C}+\frac{CG''(C)}{2}\right)\sum_{k=1}^6{\widetilde{C}_k^2}-6G(C)\mathcal{A}\right].
\end{equation}
Furthermore, since we have
\begin{equation}
\label{sum1}
\begin{aligned}
    \sum_{k=1}^{6}\widetilde{C}_k&=12\left(\frac{\partial^2C}{\partial x^2}+\frac{\partial^2C}{\partial y^2}\right)\frac{\Delta^2}{4}+\mathcal{O}(\Delta^3),\\
    &=12\mathcal{A}+\mathcal{O}(\Delta^3),
\end{aligned}
\end{equation}
and
\begin{equation}
\label{sum2}
\begin{aligned}
    \sum_{k=1}^{6}{\widetilde{C}_k}^2&=12\left(\left(\frac{\partial C}{\partial x}\right)^2+\left(\frac{\partial C}{\partial y}\right)^2\right)\frac{\Delta^2}{4}+\mathcal{O}(\Delta^3),\\
    &=12\mathcal{B}+\mathcal{O}(\Delta^3),
\end{aligned}
\end{equation}
Equation \eqref{first_3} becomes
\begin{equation}
    \label{become1}
    MCG(C)+M\left(2CG'(C)-G(C)+\frac{2G(C)}{1-C}\right)\mathcal{A}+M\left({CG''(C)}+\frac{2G(C)}{(1-C)^2}+\dfrac{2G'(C)}{1-C}\right)\mathcal{B}+\mathcal{O}(\Delta^3). 
\end{equation}
Remind that the second term in \eqref{delta1} is
\begin{equation}
    \label{become2}
    \begin{aligned}
    MCG(\newbar{K}_{\mathbf{s}}^{(\textls{m})})&=MCG(C)+MCG'(C)\widetilde{C},
    \\
    &=MCG(C)+MCG'(C)\mathcal{A}+\mathcal{O}(\Delta^3).
    \end{aligned}
\end{equation}
Then combining \eqref{become1} and \eqref{become2} gives
\begin{equation}
    \label{become3}
    \begin{aligned}
        \delta(C_{\mathbf{s}})&=\left(CG'(C)-G(C)+\frac{2G(C)}{1-C}\right)M\mathcal{A}
    +\left({CG''(C)}+\frac{2G(C)}{(1-C)^2}+\dfrac{2G'(C)}{1-C}\right)M\mathcal{B}+\mathcal{O}(\Delta^3),\\
    &=\left(CG'(C)+\frac{1+C}{1-C}G(C)\right)M\mathcal{A}
    +\left({CG''(C)}+\frac{2G(C)}{(1-C)^2}+\dfrac{2G'(C)}{1-C}\right)M\mathcal{B}+\mathcal{O}(\Delta^3).
    \end{aligned}
\end{equation}
Dividing both sides of the resulting expression by $\tau$, and letting $\Delta\to0$ and $\tau\to0$ jointly, with the ratio $\Delta^2/\tau$ held constant, leads to the following nonlinear reaction-diffusion equation,
\begin{equation}
    \label{become4}
    \frac{\partial C}{\partial t}=\nabla\cdot\left[D_0\left(CG'(C)+\frac{1+C}{1-C}G(C)\right)\nabla C\right]+\lambda CF(C),  
\end{equation}
where
\begin{equation}
\label{limitparameter1_sup}
    D_0=\frac{M}{4}\lim_{\Delta,\tau\to0}
    \frac{\Delta^2}{\tau},\quad \lambda=\lim_{\tau\to0}\frac{P}{\tau}.
\end{equation}
If we define
\begin{equation}
    D(C)=D_0\left[CG'(C)+\frac{1+C}{1-C}G(C)\right],
\end{equation}
then the continuum limit is written as
\begin{equation}
    \label{PDE_general_2D_sup}
    \frac{\partial C}{\partial t}=\nabla\cdot\left[D(C)\nabla C\right]+\lambda CF(C).
\end{equation}

\newpage
\section{Numerical method}
To numerically calculate solutions of the reaction-diffusion equation
\begin{equation}
    \label{PDE_general_2D_SupNum}
    \frac{\partial C}{\partial t}=\frac{\partial }{\partial x}\left(D(C)\frac{\partial C}{\partial x}\right)+R(C),
\end{equation}
on $0<x<L$, we first discretise the spatial derivative in Equation \eqref{PDE_general_2D_SupNum} with an $(I+1)$ uniformly distributed nodes with spacing $\delta x>0$, which are indexed by $x_i$ with $i=0,1,2,...,I$ satisfying $I=L/\delta x$. We leave the time derivative continuous and obtain
\begin{equation}
    \label{methodofline}
    \begin{aligned}
    \frac{\textls{d}C_{i}}{\textls{d}t}=&\dfrac{1}{2\delta x^2}(D(C_{i+1})+D(C_{i}))C_{i+1}\\
    &-\dfrac{1}{2\delta x^2}(D(C_{i+1})+2D(C_{i})+D(C_{i-1}))C_{i}\\
    &+\dfrac{1}{2\delta x^2}(D(C_{i})+D(C_{i-1}))C_{i-1}\\
    &+R(C_{i}).
    \end{aligned}
\end{equation}
This equation is valid for interior nodes, and is modified on the boundary nodes to simulate periodic boundary conditions. This system of $I$ coupled ordinary differential equations is then integrated through time using MATLABs function ode45 \citep{ode45ref}.

\newpage
\section{Discrete-continuum comparisons with power-law diffusivity}
Figure~\ref{fig:columnl_5} show the continuum and discrete results obtained with the nonlinear diffusivity functions $D(C)=D_0C^m$ for $m=1,2,3$ and the corresponding movement crowding functions. Discrete simulations are performed with parameters $M=1$ and $P=1/1000$, leading to $D_0=1/4$ and $\lambda=1/1000$ in the continuum model. For the discrete model, we average the simulation data over 40 times of identically-prepared realisations, and initially locate agents in the middle of the domain with a vertical strip with width $w=20$, which means that the initial condition for the continuum model is $C(x,0)=1$ for $x\in[40,60]$ and $C(x,0)=0$ elsewhere. We compare $C(x,t)$ with $\left<C(x,t)\right>$ in Figures~\ref{fig:columnl_5}(a)--(c), and compare $\mathcal{C}(t)$ with $\left<C(t)\right>$ in Figures~\ref{fig:columnl_5}(d)--(f). Results in Figure~\ref{fig:columnl_5} indicate that the match between the solutions of continuum model and the appropriately averaged data from discrete model is excellent.

\renewcommand{\thefigure}{S1}
\begin{figure}[ht]
\centering
\includegraphics[width=\textwidth]{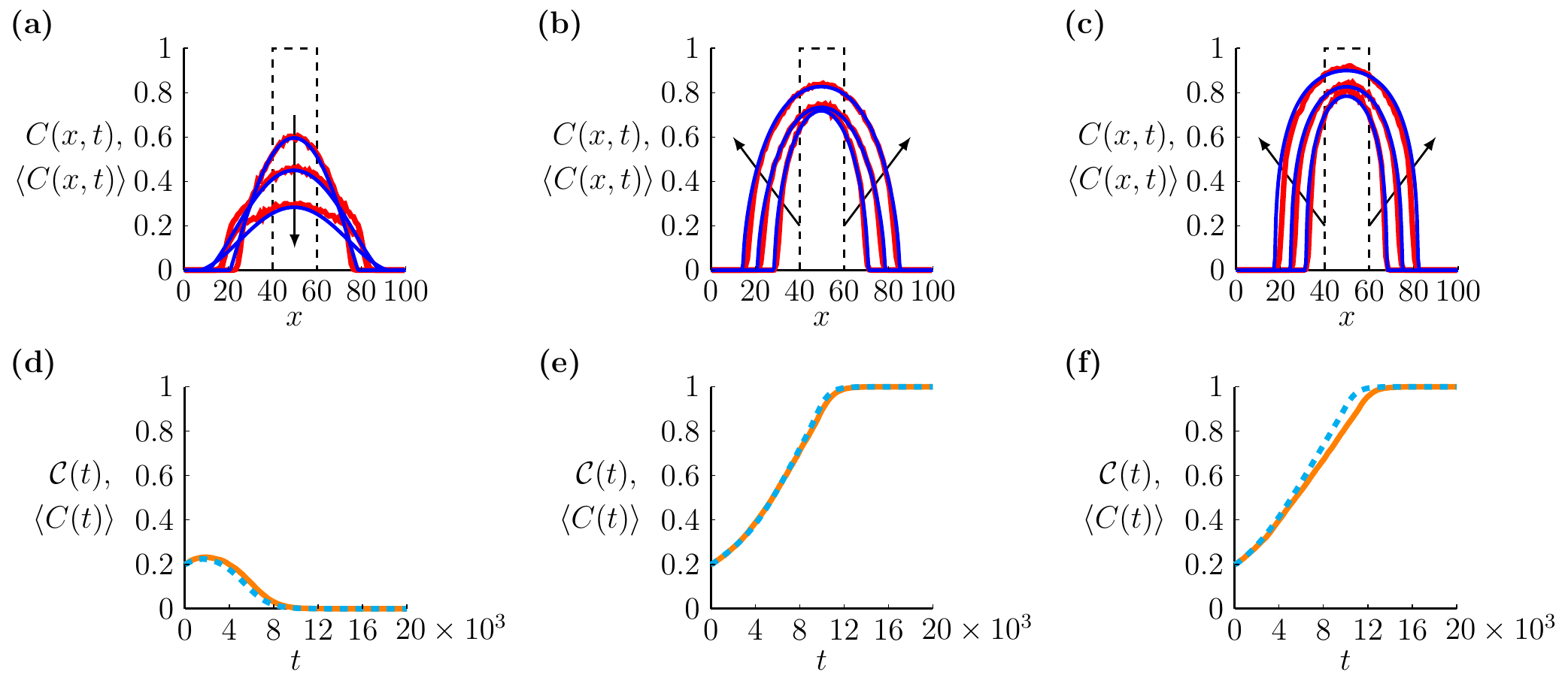}
\caption{\textbf{Population dynamics with different nonlinear diffusivity functions.} (a) $C(x,t)$ (blue) obtained with $D(C)=D_0C$, $D(C)=D_0C^2$ and $D(C)=D_0C^3$, respectively, and $\left<C(x,t)\right>$ (red) obtained with the corresponding $G(C)$, at $t=1000, 3000, 5000$. The dashed lines in (a)--(c) indicate the initial distribution, where $C(x,0)=1$ for $x\in[40,60]$ and $C(x,0)=0$ elsewhere. The arrows in (a)--(c) show the direction of increasing time. (d)--(f) $\mathcal{C}(t)$ (dashed cyan) and $\left<C(t)\right>$ (solid orange) obtained with $D(C)=D_0C$, $D(C)=D_0C^2$, $D(C)=D_0C^3$, respectively. Note that we consider $M=1$ and $P=1/1000$ leading to $D_0=1/4$ and $\lambda=1/1000$.}
\label{fig:columnl_5}
\end{figure}

\newpage
\section{Phase diagrams with the well-mixed or vertical strip initial distributions}
We show the phase diagram for the survival and extinction of the well-mixed populations with $G(C)=1-C$, $G(C)=(1-C)(1+C/2)$ and $G(C)=(1-C)(1-C/2)$ in Figure~\ref{fig:phasediagram_dis_2}(a)--(c), respectively. In our discrete simulations we randomly distribute a fixed number of agents on an $L\times L$ domain where $L=100$, so that the density is, on average, $B$ at any site. We consider the $(B, P/M)$ phase space and discretise it into $51\times40$ nodes, where we change $B\in[0.1, 0.6]$ and change $P/M\in[1/1000, 4/100]$ by varying $P$ where $M=1$ in the discrete model, and varying $\lambda=P$ where $D_0=1/4$ in the continuum model. Since different identically-prepared realisations of the stochastic model can lead to either survival or extinction of the population, for each pair of parameters $w$ and $P/M$, we generate $40$ identically-prepared realisations and compute the survival probability, $S\in[0,1]$, as the fraction of realisations in which the population survives after a sufficiently long period of time $\mathcal{T}$, which we take $\mathcal{T}=max(30/P,10^4)$ in practice.
The blue shading in Figures~\ref{fig:phasediagram_dis_2}(a)--(c) shows the survival probability $S$. For the well-mixed population, the continuum model simplifies to an ODE
\begin{equation}
    \label{ODE}
    \frac{\textls{d}C(t)}{\textls{d}t}=\lambda C(t)F(C(t)),
\end{equation}
where $F(C)=2.5(1-C)(C-A)$ with the Allee threshold $A=0.4$. Therefore, the boundary separating the survival and extinction of populations in the continuum model is always the vertical line $B=0.4$, associated with the Allee threshold $A=0.4$. Results in Figure~\ref{fig:phasediagram_dis_2} indicate that the three movement crowding functions lead to very similar boundaries for the survival and extinction of populations in the discrete model. This suggests that nonlinear diffusion plays the same role as linear diffusion on the fate of well-mixed populations.

\renewcommand{\thefigure}{S2}
\begin{figure}[ht]
\centering
\includegraphics[width=\textwidth]{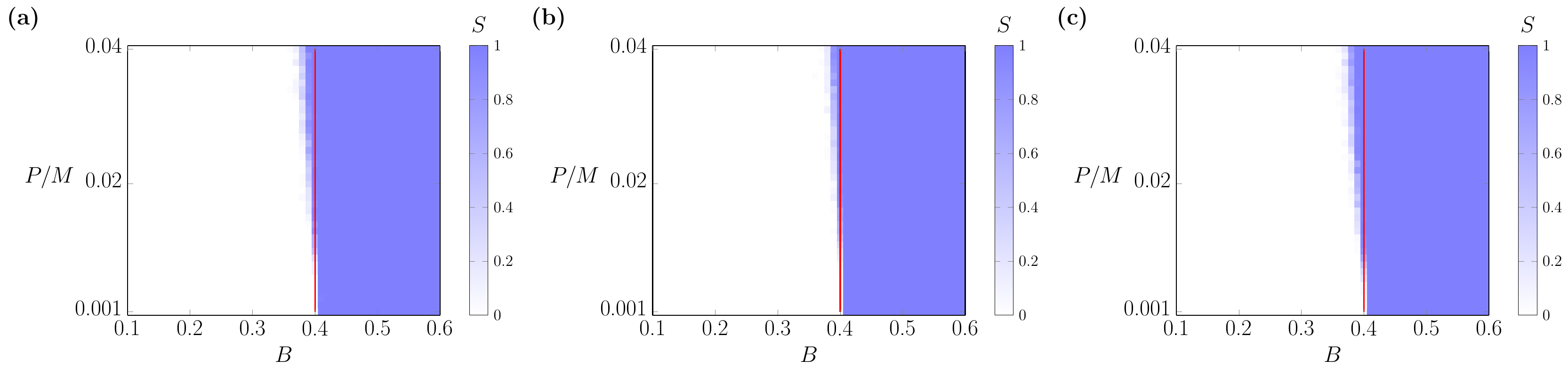}
\caption{\textbf{Phase diagrams for the survival and extinction of populations with the well-mixed initial distribution and different $G(C)$.} (a)--(c) Phase diagrams with $G(C)=1-C$, $G(C)=(1-C)(1+C/2)$ and $G(C)=(1-C)(1-C/2)$, respectively. The phase space is descretised into a rectangular mesh with $51\times 40$ nodes for $B\in[1/10,1/60]$ and $P/M\in[1/1000,4/100]$ where $M=1$. The blue shading is the survival probability in the discrete model. The red line in each phase diagram indicates $B=0.4$, which relates to the Allee threshold, $A=0.4$, and is the survival/extinction boundary for the continuum model.}
\label{fig:phasediagram_dis_2} 
\end{figure}

\newpage
We then consider the vertical strip initial distribution and compare the continuum and discrete results. We take the $(w,P/M)$ phase space and discretise it uniformly into a rectangular mesh with $41\times40$ nodes, where $w\in[0,40]$ and $P/M\in[1/1000,4/100]$. We vary $P$ and fix $M=1$ in the discrete model, leading to $\lambda=P$ and $D_0=1/4$ in the continuum model. We show the survival probability on the $(w,P/M)$ phase space in Figure~\ref{fig:phasediagram_dis_3}, and overlap with the boundary that separates survival and extinction of the population in the continuum model. We show the results obtained with the linear diffusion, increasing $D(C)$ and decreasing $D(C)$ in Figure~\ref{fig:phasediagram_dis_3}(a)--(c), respectively. The boundaries separating survival and extinction in discrete and continuum models are very close on the phase diagram, which indicates that the discrete and continuum models are consistent in generating the long-term outcomes of bistable populations.

\renewcommand{\thefigure}{S3}
\begin{figure}[ht]
\centering
\includegraphics[width=\textwidth]{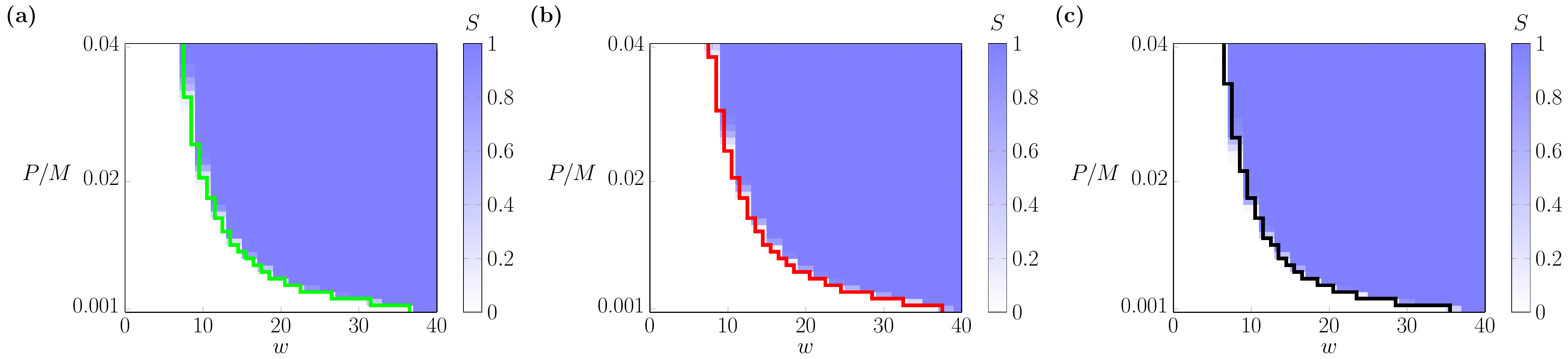}
\caption{\textbf{Phase diagrams for the survival and extinction of populations with both discrete and continuum models.} (a)--(c) Phase diagrams for the survival and extinction of populations with $G(C)=1$, $G(C)=(1-C)(1+C/2)$ and $G(C)=(1-C)(1-C/2)$, respectively. The phase space is descretised into a rectangular mesh with $41\times 40$ nodes for $w\in[0,40]$ and $P/M\in[1/1000,4/100]$ where $M=1$.  The blue shading is the survival probability in the discrete model. The solid line in each phase diagram is the survival/extinction boundary obtained from the continuum model.}
\label{fig:phasediagram_dis_3} 
\end{figure}

\newpage
\bibliographystyle{elsarticle-harv.bst} 
\bibliography{Paper4_arXiv.bib}

\end{document}